# Bacterial Motility Across Scales: Mechanisms, Live Imaging, and Quantitative Analysis

Anna Mas[1], Antoine Le Gall[1,*], Marcelo Nollmann[1,*]

[1] *Centre de Biologie Structurale, CNRS UMR 5048, INSERM U1054, Université de Montpellier, 60 rue de Navacelles, 34090, Montpellier, France*

* Corresponding authors: nollmann@cbs.cnrs.fr, antoine.legall@cbs.cnrs.fr

**Abstract |** Bacteria live in environments that are constantly changing. To survive, they rely on different motility systems that let them move, explore, and interact with their surroundings. These motility systems not only control the movement of individual cells but also give rise to collective behaviors such as coordinated spreading, cooperative predation, and multicellular development. Understanding these processes requires not only a description of the underlying molecular machines, but also quantitative observations spanning single-cell behavior and collective dynamics. Imaging approaches now make it possible to follow motility across scales, from molecular machines to bacterial communities, while computational analyses extract principles that link mechanisms to emergent dynamics. By combining molecular, behavioral, imaging, and analytical perspectives, this review provides an integrated view of bacterial motility that links single-cell behavior to community-level dynamics across scales.

**Keywords**: Bacterial motility; swimming; swarming; twitching; gliding; quantitative image analysis; live cell imaging; microscopy

## Introduction

From the earliest days of microscopy, microbial motion has fascinated scientists. In 1677, Antonie van Leeuwenhoek marveled at “many very little living animalcules, very prettily a-moving” in a drop of rainwater, revealing microbes as dynamic, self-propelled organisms rather than static forms. This raised a fundamental question: how can organisms so small move effectively? The physics of life at low Reynolds numbers, famously explored by Edward Mills Purcell, shows that in this regime inertia is negligible, drag dominates, and propulsion relies on asymmetric movement[1].

Motility is a widespread and evolutionarily conserved trait that provides a selective advantage in heterogeneous and dynamic environments. By actively moving, bacteria can navigate gradients of nutrients, pH, oxygen, temperature, and light, thereby accessing favorable conditions and avoiding harmful ones[2,3,4,5,6].

The ecological significance of motility is evident across diverse contexts, ranging from host colonization to microbial interactions in natural and multispecies communities. Pathogenic bacteria such as *Escherichia coli*, *Salmonella enterica*, *Campylobacter jejuni*, and *Helicobacter pylori* rely on it to penetrate viscous mucus and colonize epithelial surfaces[7,8,9]. In predatory bacteria such as *Myxococcus xanthus* and *Bdellovibrio bacteriovorus*, motility is essential to sustain their predatory life cycle[10,11,12]. In coculture biofilms, *Pseudomonas aeruginosa* exploits motility to overgrow *Agrobacterium tumefaciens* colonies, while the latter uses motility to emigrate and persist[13]. More broadly, motility enables bacteria both to avoid harmful conditions, such as antibiotic gradients[14,15] or microbial predators[16], and to reach competitors or preys in order to release antimicrobial molecules during interbacterial interaction[17].

These diverse ecological roles reflect the versatility of motility, which can support both individual navigation and highly coordinated behaviors. Beyond simply moving away from their colony, motility also drives social organization, for instance in *P. aeruginosa* where distinct motility systems act at successive stages of biofilm development[18], and more generally in the formation of multicellular assemblies such as fruiting bodies, mushroom-like structures, or vortice[19,20,21,22].

But how do bacteria achieve this remarkable ability to move in such diverse conditions? Over evolutionary time, they have developed distinct mechanical strategies, each with characteristic dynamics, to overcome the challenges of locomotion at the microscopic scale. Helical flagella rotate like corkscrews behind the cell, driving swimming and swarming with rotation cycles on the millisecond scale[23]. Type IV pili (T4P) grip and pull like grappling hooks, producing twitching motility through extension-retraction cycles that occur over seconds[24]. Surface gliding motors rely either on rotating conveyor belts that transport adhesins along the cell surface or on stationary adhesins that propel the bacterium forward in a corkscrew-like motion, generating movements on the scale of seconds to minute[25,26].

These systems differ in substrates and temporal dynamics, generating behaviors that span single-cell movements to collective organization, which poses specific challenges for observation and analysis. Capturing this diversity requires approaches that can resolve fast molecular events while tracking long-term population-level reorganizations. Historically, motility studies relied on manual tracking of a few cells, providing valuable insights but failing to capture variability across populations or the emergence of collective dynamics. Today, advances in live-cell imaging enable high-resolution observation of both individual trajectories and coordinated motion, while automated analysis tools allow quantitative characterization across scales. In this review, we first outline the molecular mechanisms of motility and the behaviors they generate, from individual trajectories to collective dynamics. We then focus on recent advances in live-cell imaging and computational analysis, which together enable quantitative characterization of motility across scales and in diverse environments.

# Diversity of Bacterial Motility Modes

## Flagellum-Based Motility

### *Mechanistic principles of swimming*

Swimming is one of the most widespread bacterial locomotion strategies, powered by a conserved rotary motor driven by ion fluxes[23,27,28,29]. The flagellum comprises three main elements: the basal body, which anchors the motor in the cell envelope, ensures torque transmission through concentric protein rings, and contains the flagellar type III secretion system (fT3SS) for assembly; the hook, which acts as a flexible universal joint; and the filament, a helical propeller that generates thrust[30,31,32,33]. Rapid rotation of the filament generates thrust, driving the bacterium through its environment[23,27]. For an in-depth overview of the structural organization and molecular components of the bacterial flagellum, see the comprehensive reviews by Armitage & Berry[34] and Minamino & Kinoshita[35].

### *Swimming behaviors*

Swimming performance varies widely depending on the ion gradient powering the motor. Most bacteria rely on a proton-motive force (PMF), typically established by the export of protons from the cytoplasm to the periplasm through the electron transport chain, photosynthetic reactions, or a reverse-functioning ATP synthase, thereby generating the gradient later used by the flagellar motor[36,37]. In contrast, bacteria from alkaline or hypersaline environments exploit a sodium-motive force (SMF) using modified stators, PomA and PomB, that conduct sodium ions[38,39,40]. The nature of ions strongly influences motor performance. PMF-driven motors typically rotate at ~300 Hz, whereas SMF motors can exceed 1300 Hz[37]. Swimming speeds range from ~25 µm/s in most species[28,41,42] to nearly 1 mm/s in *Ovobacter propellens*, which employs massive flagellar bundles[43].

Beyond the energetic basis provided by ion gradients, bacterial motility also depends on the regulatory capacity of flagella to switch rotational direction. Within the basal body, the C-ring, composed of FliG, FliM, and FliN, acts as a cytoplasmic platform that transmits torque and regulates rotational direction. In swimming bacteria, motors predominantly rotate counterclockwise (CCW). Directional switching is triggered when the response regulator CheY, in its phosphorylated form (CheY-P), binds to the FliM and FliN subunits of the C-ring. Once a threshold is reached, this binding induces conformational changes in FliG that alter its interaction with the stators, switching rotation to clockwise (CW)[44,45,46,47,48]. Precise control of flagellar motor rotation allows bacteria to modulate their trajectories and adopt diverse swimming strategies. These behaviors are strongly influenced by the number and spatial arrangement of flagella on the cell surface, which determine how cells navigate distinct environments.

In peritrichous bacteria, which have flagella distributed over the entire cell surface, such as *E. coli*, *Bacillus subtilis*, and *Serratia marcescens*, motility alternates between straight "runs" and brief random "tumbles" (Figure 1A,i). Runs occur when CCW motor rotation aligns the filaments into a bundle, producing linear swimming for 1-2 s. Tumbles occur when one or more motors switch to CW, disrupting the bundle and producing a brief (~0.1 s) random reorientation. This stochastic switching, regulated by the chemotaxis system, biases tumble frequency in response to cues, yielding a biased random walk that enhances nutrient foraging and avoidance of harmful conditions[41,42,49,50,51,52].

In monotrichous species (bearing a single polar flagellum) such as *P. aeruginosa* and *Azospirillum brasilense*, in amphitrichous bacteria (with a flagellum at each pole) like *C. jejuni*, and in lophotrichous species (carrying a tuft of flagella at one or both poles) including *Pseudomonas putida* and *H. pylori*, motility follows a run-reverse pattern. Because flagella are confined to one or both poles, cells reverse direction without reorienting the body: CCW motor rotation drives forward swimming, whereas CW rotation drives backward motion along the same axis (Figure 1A,ii). This strategy is particularly advantageous in structured or confined environments, where retracing paths enables gradient sensing, obstacle escape, or repeated sampling of favorable niches[53,54,55,56,57,58,59,60]. A context-dependent variant of this behavior is flagellar wrapping, where the filament coils around the cell body instead of remaining extended. Triggered immediately after reversal, this transient configuration, observed in *P. aeruginosa*, *Vibrio fischeri*, *P. putida*, *Shewanella putrefaciens*, and *C. jejuni*, facilitates reorientation or passage through tight or viscous spaces, thereby enhancing maneuverability[61,62,63,64,65].

Some monotrichous bacteria, such as *Vibrio alginolyticus* and *S. putrefaciens*, display a modified strategy known as run-reverse-flick. A forward run driven by CCW rotation is followed by a brief CW reversal, after which resumption of CCW rotation causes the flexible hook to bend sharply, producing a ~90° flick that reorients the cell (Figure 1A,iii). This behavior enables rapid reorientation and efficient space exploration, particularly in low-viscosity fluids[66,67,68].

Subpolar-flagellated bacteria such as *Rhodobacter sphaeroides* use a run-stop-run strategy, in which cells periodically halt flagellar rotation (Figure 1A,iv). During these pauses, reorientation arises not only from Brownian motion but also from the mechanical influence of the non-rotating flagellum. Anchored asymmetrically near the pole, this filament acts as a passive lever that modifies how thermal and viscous forces act on the cell body, amplifying random rotational fluctuations. This energy-efficient mechanism avoids active bundle remodeling or directional switching and is particularly advantageous in nutrient-poor or viscous environments[69,70,71,72].

For a more comprehensive overview of flagellum-driven behaviors, see the review by Grognot & Taute[73]. Motility systems based on internal periplasmic flagella, such as those found in spirochetes, are not covered in this review; for an in-depth discussion, see Nakamura[74].

### *Collective dynamics - Swarming*

Swarming is a collective mode of flagellum-driven motility that enables dense populations of bacteria to migrate rapidly across semi-solid surfaces. Unlike swimming, which relies on individual propulsion in liquid, swarming depends on morphological and behavioral adaptations that overcome surface friction and viscosity. These adaptations enhance propulsion, alignment, and coordination, enabling cells to assemble into dense, dynamic rafts that advance as a unified front while retaining flexibility to navigate complex substrates[75,76,19] (Figure 1B).

Swarming initiation integrates mechanical, chemical, and environmental cues to ensure the costly transition to collective migration occurs only under favorable conditions. Surface contact increases load on the flagellar motor, sensed as mechanosignaling that can trigger differentiation into swarmer cells and recruitment of lateral flagella or high-torque stators[77,78,79]. Quorum sensing further activates swarming once cell density is sufficient, inducing flagellar gene expression and biosurfactant secretion that lubricate the surface and reduce tension[19]. By contrast, *E. coli* and *S. enterica* bypass surfactants by modifying their lipopolysaccharides to enhance wettability[80]. Swarming efficiency also depends on environmental conditions: semi-solid substrates (0.4-1.0%) and sufficient hydration, often maintained by fluid films or biosurfactants, are essential for collective migration[81,82,83,19,84,85,86]. Nutrient responses, however, are species-specific: amino acids, carbon, phosphate, and iron can either promote or inhibit swarming depending on the organism[87,88,89,90,91,92,84,93]. Finally, swarming requires an active reduction of intracellular cyclic diguanylate (c-di-GMP), a ubiquitous bacterial second messenger that regulates transitions between motile and sessile lifestyles. This reduction is achieved by the induction of phosphodiesterases. Low c-di-GMP lifts the repression of flagellar genes and, in *P. aeruginosa*, promotes recruitment of high-torque MotCD stators essential for surface migration, whereas high levels stabilize MotAB stators optimized for swimming[94,95,96,97,98,99,100,101].

Upon swarming induction, vegetative cells undergo two major adaptations that enhance collective motility. First, they often elongate, increasing in length by 2- to 40-fold and often forming multinucleate filaments due to division arrest. In *E. coli*, *Salmonella*, and *Serratia liquefaciens*, swarmer cells can typically reach 5-20 µm, whereas in *Proteus mirabilis* they can extend up to 70 µm; in *B. subtilis*, an optimal aspect ratio maximizes efficiency, while excessive elongation impairs coordination of swarms[76,87,102,103,104]. Second, swarming triggers hyperflagellation. In *V. parahaemolyticus*, this involves the induction of lateral flagella, while in peritrichous species such as *E. coli*, *Salmonella typhimurium*, *S. liquefaciens*, and *P. mirabilis* flagellar numbers rise sharply, reaching a 20- to 40-fold increase in *P. mirabilis* with a fivefold rise in surface density[76,87,102,19,104].

In parallel, swarming involves behavioral reprogramming to maintain directional persistence and group cohesion. In *E. coli*, this is associated with a marked reduction in tumble bias, extending CCW runs at the swarm front[105]. The shift has also been observed in *P. mirabilis*, *S. marcescens*, *S. enterica*, *B. subtilis*, and *P. aeruginosa*, suggesting a conserved mechanism[106]. In some species, notably *P. aeruginosa*, swarming also involves an increased frequency of flagellar reversals, modulated by c-di-GMP, enabling sharp reorientations that can initiate new tendrils at colony edges[107].

Interestingly, the morphological and behavioral adaptations described above give rise to large-scale collective patterns. In surfactant-producing species, cells organize into dense rafts that migrate in swirling flows. The front typically forms a fast-moving monolayer, while rear cells stack into multilayered, three-dimensional dynamic clusters[19,107]. This organization drives rapid radial expansion and the formation of dendritic tendrils, branched projections that extend from the colony periphery. In *P. aeruginosa*, the architecture of these tendrils (number, width, branching) is modulated by environmental and genetic factors, highlighting the plasticity of swarming patterns[107]. Functionally, tendrils concentrate motile cells at the front to promote outward movement and surface colonization[108]. At the single-cell level, swarming increases motility compared to planktonic swimming. Speeds typically rise by 1.2- to 1.9-fold, reaching up to ~40 µm/s, reflecting increased motor output and high-torque stator recruitment[105,106].

Beyond locomotion, swarming provides major survival advantages. Swarmer cells display increased antibiotic tolerance through multiple mechanisms, including high cell density that provides physical shielding, reduced membrane permeability that limits drug uptake, and transcriptional reprogramming of stress and translation pathways[14,109,15,110]. In *E. coli*, chemotaxis-competent swarms can also escape antibiotic gradients, not by sensing the drug itself but by responding to distress signals from lysed neighbors[111,112]. This tolerance is reversible and lost upon return to the planktonic state[113,109,114,15]. Swarming cells are less sensitive to antibiotics and show transcriptional upregulation of virulence factors and secretion systems[15]. For a detailed discussion of the biology and defining features of swarming motility, see the comprehensive review by Kearns[19] or Yan et al.[115].

## Type IVa pili-Based Motility: Twitching

### *Mechanistic principles of twitching*

T4P are flexible, helical filaments composed of pilin subunits and are widespread among bacteria. They support diverse functions including adhesion, DNA uptake, biofilm formation, virulence, surface sensing, and motility[116,117]. Among their subtypes, type IVa pili (T4aP) specifically drive twitching motility, a surface-associated locomotion based on cycles of pilus extension, substrate attachment, and retraction. These dynamic events are powered by cytoplasmic ATPases and coordinated by a multi-protein complex spanning the cell envelope[24,118,119,120]. Retraction occurs at up to 1000 subunits per second (~0.8 µm/s) and generates forces above 100 pN[24,118,121,119]. While the overall architecture of T4aP systems is conserved, their molecular components vary between species. For a detailed overview of T4P structure, dynamics, and functional diversity, see the comprehensive review by Craig et al.[122] and Ellison et al.[117]. The following sections focus on the core features characterized in *Pseudomonas aeruginosa* and *M. xanthus*, two model organisms in which functionally analogous systems have been extensively studied.

### *Twitching behavior*

Twitching motility relies on the action of multiple T4aP that generate complex and sometimes antagonistic forces. This dynamic interplay enables fine directional control even on heterogeneous substrates through a “tug-of-war” mechanism[123,124].

Twitching speeds vary substantially across bacterial species. *P. aeruginosa* typically moves at around 0.15 µm/s[125], while *Neisseria* species can reach velocities up to 1-1.5 µm/s[24,126]. In contrast, *M. xanthus* moves much more slowly, averaging 2 µm/min (0.033 µm/s)[12]. The T4aP filaments that drive twitching are 5-8 nm in diameter and can extend up to 5 µm in length. Their intrinsic elasticity allows them to accommodate substantial bending and stretching during surface exploration and dynamic pilus cycles[127,118,128,129,117,130]. In *M. xanthus*, an average of $6.5 \pm 3.0$ pili per piliated pole are assembled[131], providing multiple dynamic tethers to the substrate that collectively enable traction and motility.

The mechanical action of these pili enables a rich behavioral repertoire. Typical single cell trajectories are jagged and intermittent, with frequent changes in orientation and direction. In *M. xanthus*, for instance, large fluctuations of the leading pole are observed during motion, which result from the pulling action of retracting pili[132] (Figure 2A,i). In *P. aeruginosa*, simultaneous pili activity can transiently pivot cells upright, allowing them to "walk" on their pili and explore surface topographies with increased agility[133,134] (Figure 2A,ii).

T4aP-driven motility also incorporates an important regulatory dimension: the ability to reverse direction through polarity switching. While both *P. aeruginosa* and *M. xanthus* rely on polarity switches to control directionality, they use distinct regulatory systems. The specific mechanisms governing polarity switching and reversals in *M. xanthus* will be discussed in the following section (gliding). In *P. aeruginosa*, reversals are triggered by the relocalization of the T4aP machinery to the opposite pole, a process controlled by the Chp chemosensory system[135]. Mechanical input from surface-attached pili modulates the activity of PilG and PilH, two response regulators that antagonize each other to control polarity and promote unipolar pilus extension. Through this PilG-PilH interplay, *P. aeruginosa* implements a local-excitation/global-inhibition circuit that ensures robust yet flexible control of twitching directionality[135,136]. Beyond mechanical cues, this polarity-switching machinery also supports environmental sensing. *P. aeruginosa* cells are capable of navigating chemical gradients on surfaces by coupling T4aP-mediated twitching motility with the Chp chemosensory system. This chemotactic response is achieved through a behavioral mechanism in which cells increase their reversal frequency when moving away from a chemoattractant, enabling directional correction[137].

### *Collective dynamics of twitching*

Beyond single-cell navigation, T4aP dynamics also underlie emergent forms of collective motility. In *M. xanthus*, T4aP enable social (S-) motility, a cooperative mode of surface translocation in which cells migrate in cohesive groups called swarms across soft surfaces[138,139,140]. This collective movement is reinforced by interactions between T4aP and extracellular polysaccharides secreted by neighboring cells, which serve as anchoring points for pilus retraction[141] and promote close-range adhesion, thereby fostering cohesive group migration that could enhance the efficiency of cooperative behaviors such as predation[12] (Figure 2B, i).

In *M. xanthus,* S-motility is also central to development, enabling the dense multicellular aggregation required for fruiting body formation[139,142]. Under starvation, cells aggregate into multicellular mounds that differentiate into fruiting bodies containing dormant myxospores, which can later germinate and resume vegetative growth when conditions become favorable[20,142] (Figure 2B, iii). Moreover, during both development and predation, *M. xanthus* cells self-organize into ripple patterns, which are traveling waves that promote alignment (Figure 2B, ii). Ripple formation critically depends on social motility, as T4aP-deficient mutants display severe defects in ripple formation[143,144].

In other bacteria, twitching motility also contributes to collective behaviors. For instance, in *P. aeruginosa*, during early biofilm formation, twitching cells deposit trails enriched in the exopolysaccharide Psl, which bias the trajectories of subsequent cells and promote microcolony formation[145]. . At late stages of biofilm maturation, twitching facilitates the emergence of mushroom-shaped biofilm towers: non-motile cells proliferate at fixed positions to form the stalk, while motile cells actively migrate and climb these structures to form the caps[22] (Figure 2B, iv). Although individual cells move independently, this indirect coordination via a shared extracellular environment leads to spatial self-organization and the emergence of structured communities.

## Adhesin-Based Motility: Gliding

Gliding motility is a surface-associated form of locomotion that operates without flagella or retractile pili. It relies on inner-membrane motors connected to trans-envelope systems that mobilize adhesins to generate traction. Two species serve as model organisms illustrating distinct strategies: *M. xanthus* and *Flavobacterium johnsoniae*.

### *Mechanistic principles of gliding in Myxococcus xanthus*

In *M. xanthus*, gliding motility, also referred to as adventurous (A-) motility, is powered by the Agl-Glt machinery, a PMF-driven motor embedded in the inner membrane[146]. Motor complexes are recruited at the leading pole[147], travel following a helical trajectory toward the lagging pole[26], and couple to outer membrane adhesins that form focal adhesion sites providing traction on the substrate[148]. A detailed description of the gliding machinery has recently been provided by Cambillau and Mignot[149].

### *Gliding behavior in Myxococcus xanthus*

As cells move, they rotate along their longitudinal axis[26], CW when viewed from the front, and advance at rates of ~2-4 µm/min[150,151], typically following nearly linear trajectories[132] (Figure 3A,i).

Polarity switching driven by MglA/MglB dynamics provides the mechanical basis for reversals, while the Frz chemosensory system coordinates the relocation of the MglA-MglB polarity module and controls reversal frequency[152] (Figure 3A,i) . The pathway begins with FrzCD, a cytoplasmic methyl-accepting chemotaxis protein associated with the nucleoid and lacking transmembrane domains[153]. FrzCD's activity is tuned by the methyltransferase FrzF and the methylesterase FrzG, analogous to CheR and CheB in *E. coli* [153]. FrzCD interacts with the CheW-like proteins FrzA and FrzB, of which only FrzA is essential for activating the histidine kinase FrzE, a CheA-CheY fusion homolog[153,154]. Upon activation, FrzE autophosphorylates and transfers the phosphoryl group to two response regulators: FrzX, which cooperates with RomR to relocate the MglA-MglB polarity module and reestablish the motility machineries at the new leading pole, and FrzZ, which contains two CheY-like domains and fine-tunes reversal frequency by shortening the refractory period[155,156]. On nutrient-rich agar, reversals occur approximately every six minutes and are crucial for allowing cells to bypass obstacles, maintain collective alignment, and generate multicellular patterns such as ripples and fruiting bodies[157,158].

In addition to polarity control, A-motility relies on extracellular components that mediate cell-surface interactions, most notably the extracellular matrix (ECM). *M. xanthus* cells produce and secrete ECM, primarily composed of polysaccharides and commonly referred to as "slime," although the function of slime secretion remains poorly defined[159]. Early studies suggested that slime extrusion might directly propel gliding cells, based on the observation of nozzle-like structures and slime jets[160]. However, the later discovery of focal-adhesion-based motility clarified that the ECM primarily serves as an adhesive scaffold rather than a propulsion system[26]. During active surface migration, Agl-Glt complexes direct the localized deposition of slime at discrete adhesion sites, anchoring the cell to the substratum and enabling productive movement[159]. Thus, although slime biosynthesis is independent of the motility machinery, slime deployment is tightly coupled to A-motility dynamics[159].

### *Collective dynamics of gliding in Myxococcus xanthus*

Although adventurous gliding is associated with the movement of single cells, it plays a critical role in shaping population-level organization, in part through the secretion of ECM that mediates adhesion and coordinates collective behaviors[12,161,162]. By depositing continuous slime trails, gliding cells create preferential tracks that can be followed by neighboring cells, fostering directional persistence, group migration and social cohesion[162,161,12] (Figure 3B,i). These slime trails may also serve as scaffolds for the deposition of outer membrane materials, including proteins and lipids, which could facilitate intercellular communication and the emergence of multicellular structures[163].

A-motility is also central to several multicellular behaviors in *M. xanthus*, particularly during predation. During prey invasion, gliding cells actively explore surfaces[12], penetrate prey colonies[11], and establish contact with target cells, thereby enabling contact-dependent killing and lysis through the Kil pilus system[11,164] and a type III secretion system (T3SS)-like apparatus[165]. By opening physical paths into prey communities, these gliding cells enable large groups of followers to advance, thereby amplifying the efficiency of the collective attack[11] (Figure 3B, i). Gliding motility is also essential development: on hard substrates, individual A-motile cells explore the surface and progressively converge to form multicellular clusters that initiate the developmental program. Mutants deficient in Agl-Glt components are unable to form proper fruiting bodies, highlighting the essential role of A-motility in multicellular morphogenesis on hard substrates[142,139].

### *Mechanistic principles of gliding in Flavobacterium johnsoniae*

*F. johnsoniae* glides using the Type IX Secretion System (T9SS), a PMF-driven rotary machinery embedded in the inner membrane and spanning the cell envelope. The motor drives a tread that moves along a stationary multirail scaffold beneath the outer membrane, propelling adhesins across the cell surface in defined helical trajectories. Among these, the adhesin SprB is essential: its movement along the cell body generates localized traction, causing the bacterium to rotate CCW around its long axis (when viewed from the rear) and advance with a screw-like motion[166,25,167]. For a recent structural and mechanistic overview of the Type IX secretion system underlying *F. johnsoniae* gliding motility, see the review by Trivedi et al.[168].

### *Dynamics of T9SS-based gliding*

The T9SS-driven gliding mechanism enables *F. johnsoniae* to glide at speeds up to 2 μm/s[169,170,171], typically following random orientations[21]. However, when cells remain physically connected to one another under starvation conditions, their trajectories exhibit a consistent left-turning bias that shapes the initial expansion pattern of colonies[21] (Figure 3A, ii).

Although T9SS-based gliding is an individual mode of motility, it contributes to population-level organization through environmental remodeling. Cells deposit a poorly characterized ECM that can smooth surface irregularities and create adhesive tracks guiding follower cells[21]. At the colony scale, this guidance and the directional bias in individual trajectories drive the emergence of large-scale spatial patterns, culminating under nutrient limitation in the formation of ~250 μm rotating vortices. These structures promote local alignment and sustained motion, likely enhancing colony spreading under starvation conditions[21] (Figure 3B, ii) .

## Microscopy Techniques for Live Imaging of Bacterial Motility

. Depending on the system, this requires balancing spatial and temporal resolution, imaging depth, and the invasiveness of the method. Historically, motility studies relied on manual tracking of a few cells, providing valuable insights but failing to capture variability across populations or the emergence of collective dynamics. Advances in live-cell imaging now enable high-resolution observation of both individual trajectories and coordinated motion, while automated computational tools allow quantitative characterization across scales. The choice of technique is thus guided by the type of motility under investigation, the environmental context, and the biological scale of interest, from protein localization to single-cell tracking or population-level dynamics.

### Label-free microscopy

Label-free microscopy allows the observation of live bacteria without fluorescent labeling by exploiting their intrinsic optical properties. Classical approaches based on transmission provide straightforward visualization of cells, while interferometric methods extend these possibilities to 3D tracking and nanoscale detection.

#### *Classical contrast-based methods*

Brightfield microscopy is one of the oldest and simplest optical methods[172]. It produces images from transmitted light collected by the objective lens, with image contrast and resolution adjusted primarily through the condenser and aperture diaphragms[173]. This approach enables observation over large fields of view and is useful for qualitative monitoring of bacterial populations during surface migration. However, because live bacteria are small and weakly absorbing, the contrast is poor, making cell boundaries difficult to distinguish. For clearer visualization of morphology, contrast-enhancing methods such as phase contrast, differential interference contrast (DIC), or darkfield microscopy are generally preferred.

Phase contrast microscopy converts phase shifts caused by refractive index differences into visible intensity variations, making bacteria appear darker on a light background, often surrounded by a halo[174,173]. DIC uses interferometry to generate images with enhanced edge definition[175,173]. Darkfield selectively detects scattered light, rendering cells and thin structures bright against a dark background[176,173]. While these methods greatly improve contrast compared to brightfield, they remain poorly suited for resolving individual filamentous structures such as single flagella or pili. Nonetheless, they have been used to visualize dense flagellar bundles in large spiral-shaped or marine bacteria[177,178], as well as in *S. typhimurium*[179,180].

#### *Interferometric and scattering-based methods*

Digital holographic microscopy (DHM) is an imaging technique that records both the amplitude and phase of light, enabling numerical reconstruction of cell shape and motion in 3D[181]. Instead of capturing only intensity in a single focal plane, DHM encodes an interference pattern between the object beam (light transmitted through and scattered by the sample) and a reference beams, which can be used to computationally reconstruct virtual focal stacks without physical scanning[182,183,184,185]. While not sufficient for subcellular resolution, DHM readily resolves 3D bacterial trajectories, including lateral displacements and axial fluctuations, and is particularly suited for quantitative analysis of swimming motility in liquid media, where bacteria move freely in 3D (Figure 4A, i, ii). Peng et al.[186] used DHM to investigate how substrate stiffness influences *E. coli* swimming near surfaces, showing that rigid substrates increase confinement and thereby highlighting the method's ability to resolve vertical cell positioning. In another study, Cheong et al.[187] simultaneously tracked hundreds of cells from multiple species, enabling comparative analyses of motility across genetic backgrounds and chemical conditions. Together, these studies demonstrate DHM's capacity for non-invasive, high-throughput volumetric tracking of bacterial motion.

Whereas DHM excels at volumetric tracking in bulk liquid, interferometric scattering (iSCAT) pushes the limits of label-free imaging at interfaces by detecting nanometric structures near reflective surfaces with exceptional spatial and temporal resolution[188,189,190] (Figure 4A, iii). It is particularly powerful for real-time, label-free tracking of rapid nanoscale processes at solid-liquid interfaces, as shown by Talà et al.[120], who visualized the stochastic extension and retraction cycles of individual T4P in *P. aeruginosa,* revealing the stochastic nature of pilus activity at the single-filament level (Figure 4A, iv). Given that detection sensitivity decays rapidly with distance from the reflective surface, flagella are particularly difficult to image; they can occasionally be detected when immobilized or favorably oriented[120], but their full dynamic behavior remains largely inaccessible. Overall, iSCAT occupies a unique niche for probing bacterial dynamics that are challenging to resolve with fluorescence or classical scattering methods.

## Fluorescence microscopy

While label-free techniques permit observation of bacterial behavior under near-physiological conditions, they lack molecular specificity and therefore cannot resolve the actors underlying motility. Fluorescence microscopy overcomes this limitation by specifically labeling proteins, structures, or cell populations of interest (genetically or chemically) and can be implemented through widefield, point-scanning, or light-sheet illumination approaches, each presenting distinct trade-offs in phototoxicity, recommended applications, and limitations.

### *Widefield illumination without optical sectioning*

Epifluorescence is the most widely used fluorescence mode, in which excitation and emission light share the same objective, illuminating the entire field of view simultaneously and enabling rapid imaging across large areas[191,192,173] (Figure 4B, i). It is well suited for real-time imaging of motile bacteria, enabling applications such as tracking regulatory proteins, visualizing flagella or pili, and distinguishing subpopulations within biofilms. For example, Turner et al.[51] labeled both the cell body and flagella of *E. coli* with a chemical fluorophore and used high-speed epifluorescence to precisely measure swimming speed and the rotational dynamics of the cell body and flagellar bundle (Figure 4B, ii). Kühn et al.[136] used fluorescent fusions to track the intracellular localization of the motility regulators PilG and PilH in *P. aeruginosa*, showing how they antagonistically control T4P extension via polarity regulation (Figure 4B, iii). Rombouts et al.[12] mixed two *M. xanthus* strains, each defective in one motility system (A or S) and expressing a distinct fluorescent membrane marker, and simultaneously tracked both strains during predation, revealing the spatial interplay between the two motility systems. However, epifluorescence produces substantial out-of-focus signal in thick samples[192], which can reduce contrast and obscure fine details. This limitation is mitigated by approaches such as Total Internal Reflection Fluorescence (TIRF) and Highly Inclined and Laminated Optical sheet (HILO) microscopy, which restrict excitation to thinner optical sections.

In TIRF microscopy, an oblique laser beam undergoes total internal reflection at the glass-sample interface, generating an evanescent field that penetrates only ~100-500 nm into the specimen[193,192]. This selective excitation near the substrate minimizes background and is particularly suited for studying dynamic structures at the cell-surface interface, such as focal adhesion complexes, pili or flagellar motors (Figure 4B, iv-vi). For instance, Faure et al.[26] exploited the surface selectivity of TIRF to visualize fluorescently tagged reporters of motility in *M. xanthus*, and detected Agl-Glt complexes as they transitioned from dynamic intracellular motion to stationary focal adhesions at the membrane-substrate interface. Using the same TIRF-based approach, Leake et al.[194] quantified the number of MotB subunits fused to fluorescent proteins within active flagellar motors by stepwise photobleaching, and further showed that these subunits undergo dynamic turnover in living cells (Figure 4B, vii).

In HILO microscopy, a laser beam is inclined just below the critical angle for total internal reflection, providing a compromise between epifluorescence and TIRF[195] (Figure 4B, viii). This illumination reduces out-of-focus fluorescence, yielding a higher signal-to-noise ratio for motility reporters and surface-associated structures such as pili (Figure 4B, ix, x). For example, Koch et al.[196] used HILO microscopy to track T4P in *P. aeruginosa* labeled using a chemical labeling, capturing in real time their stochastic cycles of extension, pausing, and retraction, which occurred at random intervals within single cells. The reduced excitation volume of HILO allowed detection of individual pili, revealing dynamic events that would be obscured under conventional epifluorescence.

### *Point-scanning with optical sectioning*

Confocal laser scanning microscopy (CLSM) improves image contrast by restricting fluorescence excitation and detection to a confined focal volume defined by a pinhole, thereby enabling high-contrast optical sectioning and 3D reconstruction of thick samples[191,192] (Figure 4C, i). Its point-by-point scanning is relatively slow and can be phototoxic, making it less suitable for capturing fast processes occurring on millisecond-to-second timescales in live cells. These trade-offs were evident in the study by Ravel et al.[197], who tracked *Bacillus* trajectories within a *Staphylococcus aureus* biofilm, with each species labeled by a distinct fluorophore (Figure 4C, ii). To preserve temporal resolution, they used two-dimensional (2D) time-lapse imaging, acquiring a single XY plane at each time point, since volumetric CLSM scanning was too slow to capture the required dynamics. This case underscores the practical limitations of confocal imaging for fast 3D live-cell tracking.

To address these limitations, spinning disk confocal microscopy parallelizes excitation through a rotating Nipkow disk with thousands of pinholes, enabling high-speed optical sectioning with reduced phototoxicity[198,199,192] (Figure 4C, iii). This makes it well suited for live 3D imaging of fast cellular processes. For example, Moriarty et al.[200] tracked fluorescently labeled *Borrelia burgdorferi* in living mice using spinning disk microscopy, capturing both 2D and 3D trajectories of spirochetes escaping from postcapillary venules and migrating within the ear vasculature (Figure 4C, iv, v). In scattering specimens such as bacterial biofilms, the imaging depth of spinning disk confocal microscopy is limited to a few tens of micrometers, as out-of-focus fluorescence and pinhole crosstalk degrade contrast in dense samples[199,192]. Despite this, spinning disk confocal remains a powerful tool for high-speed imaging of motile bacteria *in vivo*.

### *Light-sheet illumination with optical sectioning*

Optical sectioning can also be achieved by confining excitation to a thin plane within the sample, as in light-sheet microscopy, which reduces out-of-focus illumination, photobleaching, and phototoxicity compared to widefield or confocal imaging[201,202,203] (Figure 4D, i). This configuration enables rapid 3D imaging of large or thick samples, such as biofilms, with minimal damage, making it possible to follow bacterial populations in 3D over extended periods. For example, Khong et al.[204] used light-sheet microscopy to study the swimming behavior of *P. aeruginosa* stained during early stages of biofilm formation near a vertical surface. By labeling bacteria with a DNA-intercalating dye and acquiring z-stacks with 0.46 µm axial steps, they reconstructed full three-dimensional trajectories and quantified depth-resolved motility (Figure 4D, ii). This illustrates how light-sheet microscopy combines high speed, low phototoxicity, and volumetric imaging for investigating bacterial behavior near interfaces.

While traditional Selective Plane Illumination Microscopy (SPIM) configurations use two objectives and impose geometric constraints[201,192,203], , more recent designs have explored single-objective geometries and advanced illumination strategies. Oblique Plane Microscopy (OPM) builds on the principle of remote focusing introduced by Botcherby et al.[205], enabling imaging along an oblique focal plane while correcting aberrations; in OPM, both oblique light-sheet illumination and fluorescence detection occur through the same objective lens[206]. Swept Confocally Aligned Planar Excitation (SCAPE) microscopy represents a related implementation, using a tilted light sheet scanned by fast mirrors through a single objective to achieve rapid volumetric imaging[207]. In contrast, lattice light-sheet microscopy (LLSM) retains a dual-objective configuration, with excitation shaped into a lattice of Bessel beams by an objective positioned at right angles to the emission-collection objective. LLSM improves resolution and illumination uniformity while maintaining low phototoxicity[208]. Although OPM, SCAPE, and LLSM are conceptually well suited to probe bacterial motility in complex environments such as biofilms or host tissues, their application to bacterial motility remains limited.

Taken together, label-free and fluorescence-based microscopies provide complementary methods to image bacterial motility, from global population behavior to the dynamics of individual molecular machines. Their combined use has become essential for linking cellular movement with the underlying structural and regulatory mechanisms. A comparative overview of the main imaging modalities discussed above, including their phototoxicity, recommended applications, and limitations, is provided in Table 1.

# Image Data analysis

## Segmentation strategies

Segmentation is a key step for extracting quantitative information from microscopy images. It assigns pixels to discrete objects, typically bacterial cells, producing binary or labeled masks whose shape, size, position and spatial distribution can be derived. These masks underpin downstream analyses such as tracking and motility quantification.

Traditional segmentation methods relied on local features such as intensity or gradients. Thresholding (e.g., Otsu[209]), edge detection with contour tracing algorithms (e.g., Canny[210], Suzuki-Abe[211]), and watershed algorithms are effective under good signal-to-noise but often fail in crowded, noisy, or unevenly illuminated bacterial images. Morphological operations (e.g., dilation, erosion) can improve results but require manual tuning and remain difficult to generalize across large or diverse datasets. Nevertheless, such approaches have been widely used for bacterial motility studies until recently. For example, Faure et al.[26] segmented *M. xanthus* cells in kymographs using an iterative edge-detection algorithm based on the Canny method, followed by manual correction when cell contours were ambiguous. As reviewed by Jeckel & Drescher[212], many user-friendly analysis tools for bacterial cell biology and microbial communities still rely on such classical methods, typically with adjustable parameters that allow researchers to adapt segmentation to diverse imaging conditions.

Deep learning-based methods have markedly advanced segmentation in challenging conditions. Convolutional neural networks[213,214] (CNNs) learn spatial features directly from raw images, which makes them more robust to variable backgrounds, crowding and sparse signals. Although most models require annotated datasets for training, they can generalize across imaging conditions once trained. Numerous deep learning tools for bacterial segmentation have been proposed, and Jeckel & Drescher[212] provide an overview of these approaches. MiSiC, for example, uses a U-Net architecture, a type of CNN designed for pixel-wise image segmentation[213], trained on annotated and synthetic bacterial images, enabling robust segmentation across brightfield, phase contrast and low-contrast fluorescence[215]. Omnipose extends the Cellpose framework with vector flow predictions to segment tightly packed or irregularly shaped bacterial cells in both label-free and fluorescence modalities[216] (Figure 5A,B). These tools provide pretrained weights and allow fine-tuning on user data. Other platforms, such as RABiTPy[217], integrate classical and deep learning pipelines optimized for bacterial datasets. Modular strategies combining multiple models and modalities have also been proposed; for instance, Rombouts et al.[218] segmented mixed *M. xanthus* and *E. coli* populations by merging predictions from five independent U-Nets trained on brightfield and fluorescence images into a confidence-weighted output. Such approaches demand substantial computational resources, careful preparation of training data and their generalizability is constrained by strain morphology, labeling strategy and imaging conditions. For users who prefer an interactive workflow, tools like Ilastik[219] remain valuable. By training machine learning classifiers on user-labeled pixels and local features such as intensity, gradient or texture, Ilastik is well suited for atypical imaging setups, small datasets or rapid prototyping, and is often used to generate training labels for deep networks.

Regardless of the method, post-processing steps such as size or shape filtering, artifact removal and manual correction of fused or fragmented masks remain essential. In motility studies, segmentation errors directly propagate into tracking and distort trajectories or dynamics, highlighting the importance of mask validation.

## Tracking

Tracking is a central component of time-lapse image analysis. It enables the reconstruction of object trajectories, such as cells or fluorescently labeled protein complexes, across sequential frames. These trajectories provide the basis for quantifying motion, interactions and dynamic behaviors in bacterial populations. Tracking complexity depends on object density, motion type, segmentation quality, and the spatial and temporal resolution of imaging.

For cell tracking, segmented masks are linked between frames using different strategies. Classical tracking methods rely on geometric rules. Nearest-neighbor linking is effective in sparse scenes with smooth trajectories but fails in dense environments, where cells may switch places or overlap[220]. Mask-overlap strategies link cells between frames by comparing the overlap of their segmented masks; the object in the next frame that shares the largest fraction of area with the previous mask is considered the same cell[221]. These approaches better tolerate morphological changes, yet remain sensitive to segmentation errors such as fused masks or ambiguous contacts[221]. More robust approaches treat linking as a global optimization problem, where all possible links between objects across frames are considered simultaneously rather than step by step. This can be formulated as a linear assignment problem, in which a cost matrix (e.g. distances or intensity differences) is constructed between objects in consecutive frames, and the optimal solution that minimizes the total cost can be efficiently obtained with the Hungarian algorithm[222,220].

In more challenging contexts, such as high cell density, low signal-to-noise, or irregular trajectories, predictive and global optimization methods offer superior performance. Kalman filters, for example, model each trajectory as a stochastic process with predictions updated by observed positions[223]. They accommodate temporary detection failures and measurement noise, but require accurate tuning and assume relatively smooth motion. Graph-based approaches generalize tracking further: detections across time are treated as nodes connected by edges weighted by spatial and temporal similarity, and trajectories are reconstructed by finding optimal paths through the graph[224]. These methods naturally account for events such as merges, splits, and missing data, though they are more computationally intensive. Hybrid strategies adapt these concepts to bacterial tracking; for example, the "bacto_tracker" framework[218] links segmented cells using proximity and shape features, then resolves ambiguous associations using a two-pass decision process based on overlap, area, and cell length (Figure 5C). While not a full graph-based or probabilistic tracker, it illustrates how custom rules can increase robustness in dense or mixed-species datasets. Several reviews summarize available software for cell tracking[212,225]. Commonly used tools include TrackMate[226,227], u-track[220], Laptrack[228], or Trackpy[229], which implement modular pipelines combining detection, frame-to-frame linking, and gap closing. These frameworks offer varying levels of automation and flexibility, enabling reliable trajectory reconstruction across a broad range of experimental conditions.

Tracking fluorescent foci, such as motility regulators or signaling complexes, relies on dedicated spot detection rather than segmentation because these structures are diffraction-limited, sparse, and dynamic. Briefly, classical approaches like Laplacian of Gaussian (LoG) filtering, implemented in tools such as TrackMate[226,227], identify local maxima but are sensitive to background heterogeneity and densely packed spots. More recently, deep learning-based detectors such as deepBlink[230] or SpotiFlow[231] provide more robust and precise spot localization than conventional LoG filters, particularly in noisy or heterogeneous images. Once detected, spots can be linked using nearest-neighbor rules, cost minimization (LAP/Hungarian), or graph-based approaches, as described above for cell tracking (Figure 5D).

## Trajectory-based quantification

Single-cell trajectories can be transformed into quantitative descriptors spanning different scales. Local kinematic features such as speed, acceleration and angular velocity capture instantaneous dynamics, while geometric properties like curvature, tortuosity and persistence describe path structure. Global statistics, including net displacement, mean squared displacement (MSD) and radius of gyration, quantify exploration, diffusivity or confinement. Higher-order measures such as entropy, turning angles and collinearity indices provide additional insights into randomness, bias and collective alignment. At the subcellular level, fluorescence tracking of labeled foci enables extraction of complementary readouts such as dwell time, localization stability, or directionality.

Trajectory quantification was used to produce multiple biological insights . In *Bacillus spp.*, quantitative analysis of cell tracks revealed divergent confinement responses when cells encountered *S. aureus* biofilms, highlighting species-specific adaptations to heterogeneous environments[197]. In *Magnetospirillum gryphiswaldense*, measurements of trajectory curvature and alignment allowed disentangling the relative contributions of magnetic torque and wall interactions to navigation[232]. Similarly, trajectory-based classification in *Proteus mirabilis* colonies uncovered distinct motility modes at the colony edge, suggesting the coexistence of two behavioral regimes within expanding swarms[233].

Beyond global descriptors, morphodynamic analysis can resolve temporal structure within trajectories by segmenting them into behavioral phases such as runs, pauses or reversals. This enables quantification of switching rates, dwell times or reversal frequencies, which has been applied to chemotaxis, surface sensing and regulatory control of motility[234,235,236,237,238]. Such behavioral transitions can also be detected automatically using statistical approaches such as change-point analysis or Hidden Markov models, as demonstrated by Ravel et al.[197].

While the descriptors outlined above can be computed with custom scripts, several general-purpose tools now provide built-in modules for motility quantification. TrackMate[226,227] offers modular calculation of frame-by-frame and aggregate metrics, while RABiTPy[217]supports batch analysis of bacterial trajectories with integrated filtering and visualization. FAST[239] is optimized to densely packed collectives and enables temporal classification of motility behaviors. Together, these tools provide scalable and standardized extraction of motility metrics across large datasets.

## Collective analyses and interaction patterns

While single-cell trajectories reveal individual strategies, many bacterial behaviors arise from population-level coordination. Swarms, colonies and biofilms exhibit spatial organization shaped by physical interactions, signaling and environmental constraints. Capturing collective phenomena such as vortices, streaming or clustering often requires going beyond individual trajectories, either by quantifying correlations between them or by using dense motion estimation methods.

### *Object-Based Tracking and Coordination Metrics*

At the local scale, when cell tracking is available, velocity vectors derived from single trajectories allow quantification of coordination between neighboring cells. Metrics such as velocity correlation, angular dispersion and polar order parameters capture short-range synchrony[240,241,242,243,244,245,246,247;248,249]. At larger scales, individual velocities can be coarse-grained into spatial vector fields using binning or kernel averaging. This enables the calculation of higher-order descriptors such as vorticity, divergence and global alignment indices, which reveal collective patterns including jets, streaming and rotational flows[240,241,242,243,244,245,246,247,248,250,251,249].

### *Clustering Analyses of Collective Motility*

Clustering algorithms can be applied to trajectory data to detect groups of bacteria moving in close proximity over time (Figure 5E). These methods reveal transient or stable motile assemblies without requiring predefined labels. For instance, Jeckel et al.[246] analyzed *B. subtilis* swarms using spatial proximity criteria to reveal collective structures emerging in different swarming phases. Be'er et al.[247] similarly defined clusters based on intercellular distances and relative speeds, uncovering a phase diagram of dynamical regimes controlled by cell shape and density. Standard clustering techniques, including k-means, hierarchical clustering or density-based approaches such as DBSCAN[252], can be applied directly to cell positions, velocities, or trajectory-derived features to segment populations into groups with similar spatial proximity or motion patterns. Such analyses help distinguish structured collective behaviors from stochastic fluctuations and highlight the conditions favoring local coordination in bacterial populations.

### *Tracking-Free Motion Estimation: Optical Flow and Particle Image Velocimetry*

In high-density bacterial populations or low-contrast images where segmentation is unreliable, dense motion estimation methods offer an alternative. Optical flow algorithms, such as Horn-Schunck and Farnebäck, compute pixel-wise velocity fields by tracking local intensity changes between frames[253]. These methods provide high spatial resolution but can be sensitive to noise and require regularization. While recent deep learning-based optical flow models such as FlowNet[254] or RAFT[255] have gained traction in computer vision, their application to bacterial motility remains largely unexplored and is not yet validated in this context. Here, we implemented RAFT on probability maps generated from time-lapse images of *M. xanthus* using the segmentation pipeline of Rombouts et al.[218], and found that it accurately captured local motion fields and directional patterns within groups of cells (Figure 5F). Particle Image Velocimetry (PIV) estimates regional motion by cross-correlating subimages between frames, yielding coarser but more robust flow fields. Compared to optical flow, PIV is less sensitive to noise and is well suited for capturing large-scale flow organization in dense bacterial communities.

Optical flow and PIV have been applied to capture regional motion patterns, enabling the quantification of local deformations, the detection of large-scale flow organization, and the mapping of spatiotemporal dynamics across entire bacterial colonies. From these dense velocity fields, researchers can compute deformation rates, identify alignment structures, and detect transitions between collective states[240,241,244,243,249,251,245,233,256].

## Conclusion and Outlook

Bacterial motility exemplifies a phenomenon that spans molecular machines, individual cells, and emergent collective behaviors. While recent advances have illuminated each of these levels, the central challenge remains to understand how activity at one scale propagates to the next, shaping adaptive strategies and ecological outcomes.

New imaging and analytical approaches are beginning to bridge these scales. High-resolution microscopy now captures the dynamics of motors and pili with molecular precision, while population-level methods quantify flows, alignments, and collective transitions. Crucially, the integration of these perspectives reveals not only how bacteria move, but also how physical constraints and regulatory programs interact to produce complex spatial organizations.

Looking ahead, the field is poised to benefit from the convergence of high-throughput microscopy, real-time adaptive imaging, and automated analysis pipelines. Yet challenges remain. Computational tools must be both generalizable across systems and sensitive to subtle motility phenotypes. Standardized benchmarking datasets, robust validation protocols, and open-source frameworks will be critical for progress. At the same time, incorporating theoretical modeling and simulations explicitly tied to empirical data will be essential to move from description toward prediction, enabling frameworks that explain and forecast bacterial behavior in complex environments.

Ultimately, the study of bacterial motility is entering a new phase: one in which multiscale observations are not just possible, but indispensable for decoding how bacteria move, adapt, and organize. As experimental and computational tools continue to evolve, so too will our capacity to link molecular mechanisms to multicellular dynamics, revealing how simple organisms navigate the physical world with remarkable precision and versatility, and opening new avenues to control, harness, or disrupt microbial movement in health, industry, and the environment.

**A** Swimming behaviors

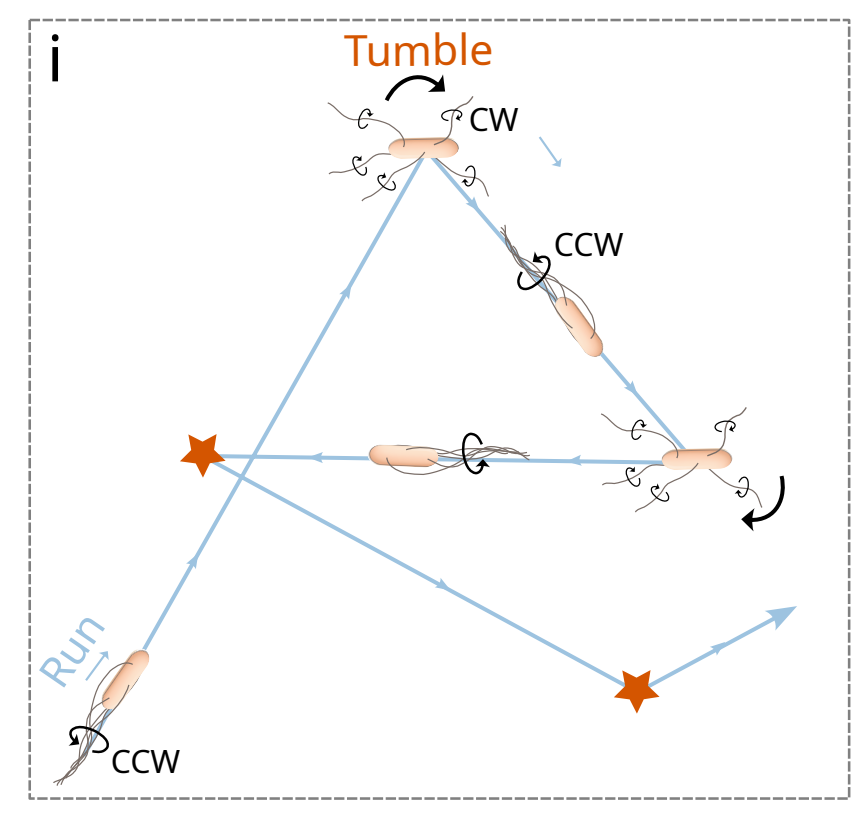

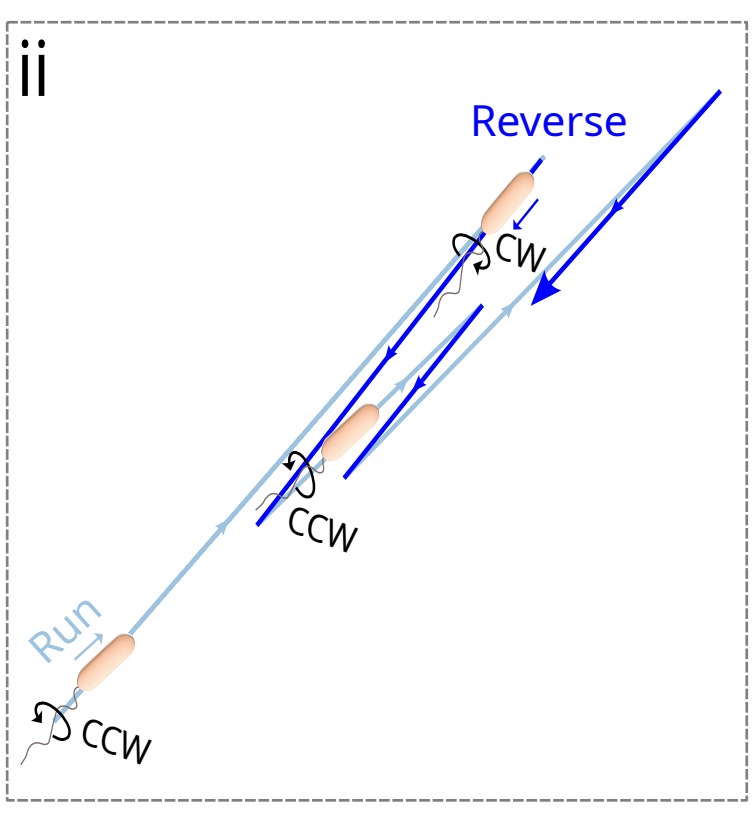

**B** Collective dynamics - Swarming

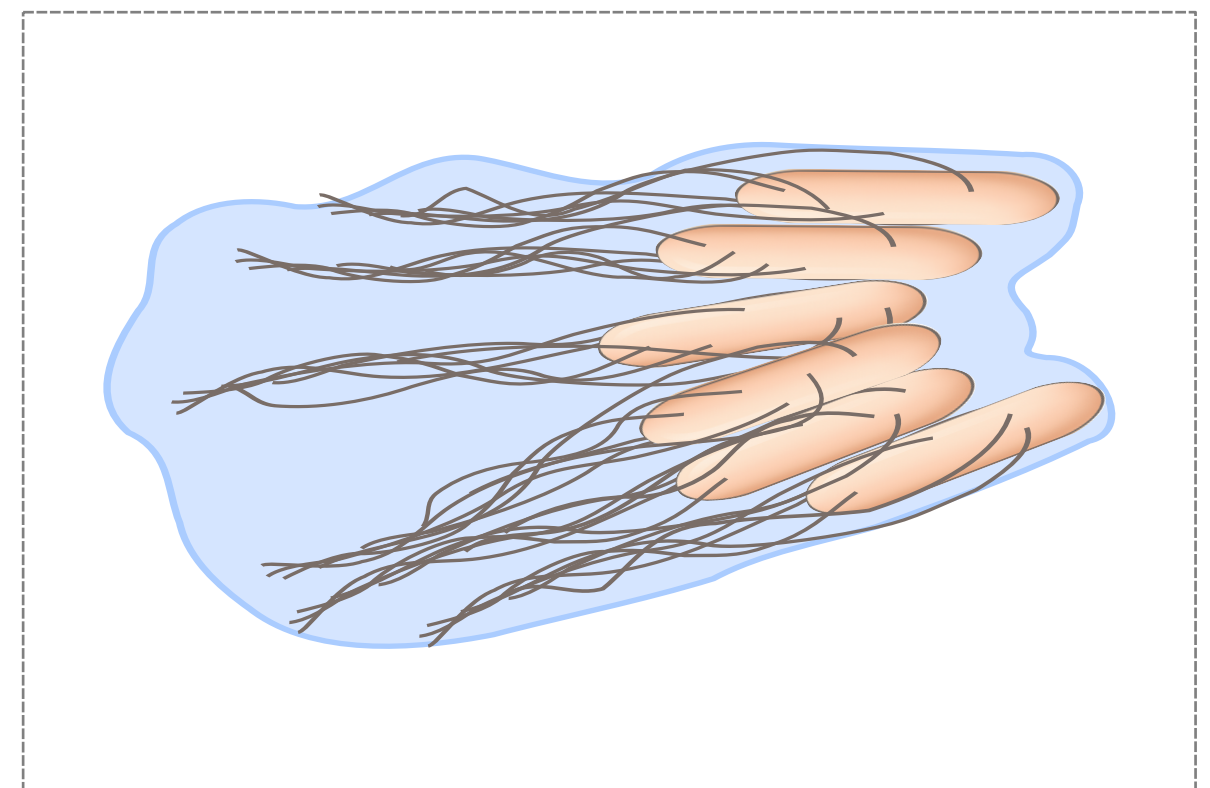

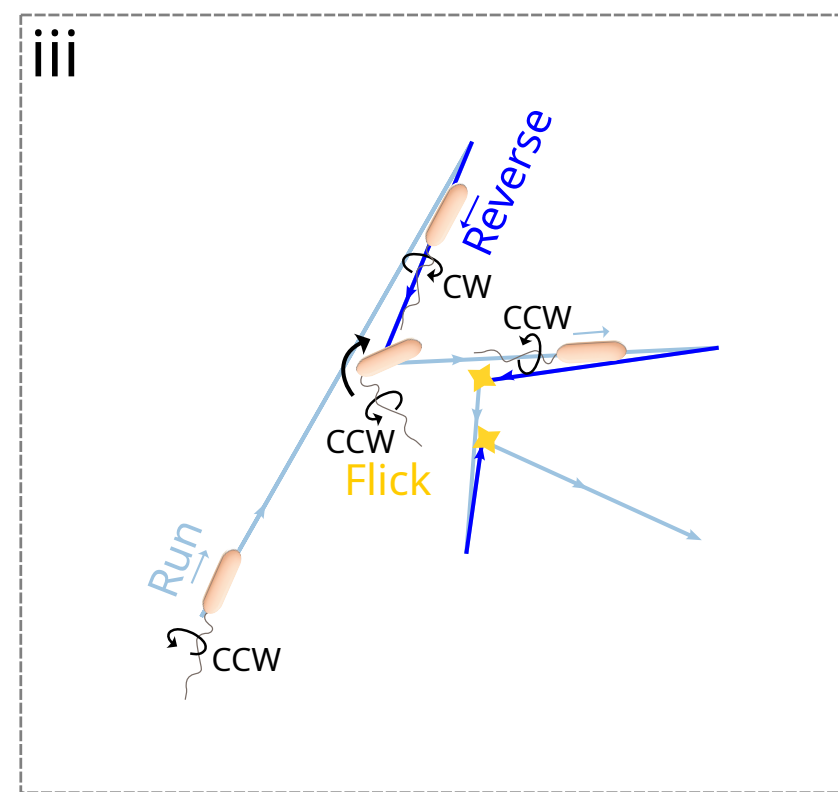

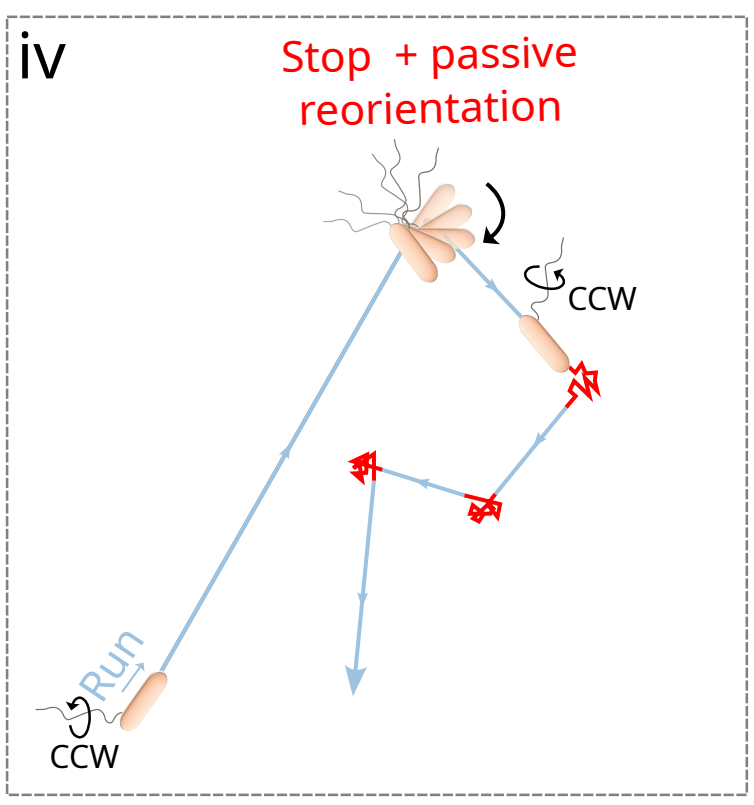

**Figure 1: Flagellum-Based Motility**

**A) Swimming behaviors.** Examples of flagellum-driven swimming strategies observed across bacteria. (i) Run-Tumble: counterclockwise (CCW) rotation bundles filaments for straight runs; clockwise (CW) rotation of one or more motors disrupts the bundle, producing brief tumbles and random reorientation. (ii) Run-Reverse: reversal of rotation direction from CCW to CW causes backward swimming along the same axis. (iii) Run-Reverse-Flick: after a short reverse, CCW rotation resumes and a hook-induced flick reorients the cell by ~90°. (iv) Run-Stop-Run: transient pauses in rotation allow passive reorientation of the cell.

**B) Collective dynamics - Swarming.** Schematic representation of the swarmer cell phenotype and its contribution to collective motility. Swarmer cell differentiation at increased cell density involves cell elongation, hyperflagellation, and surfactant production, promoting cell alignment, collective cohesion, and rapid surface colonization.

**A** **Twitching behaviors**

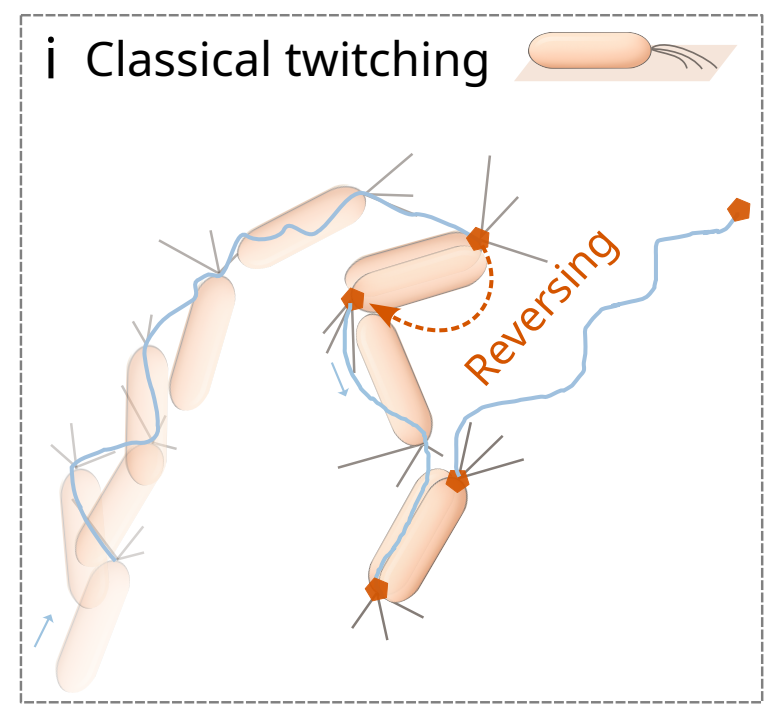

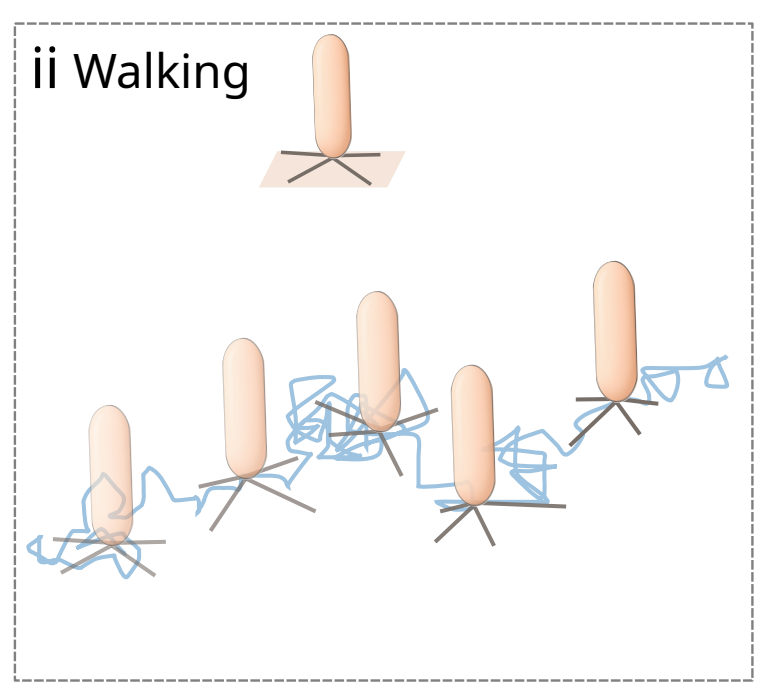

**B** **Collective dynamics of twitching**

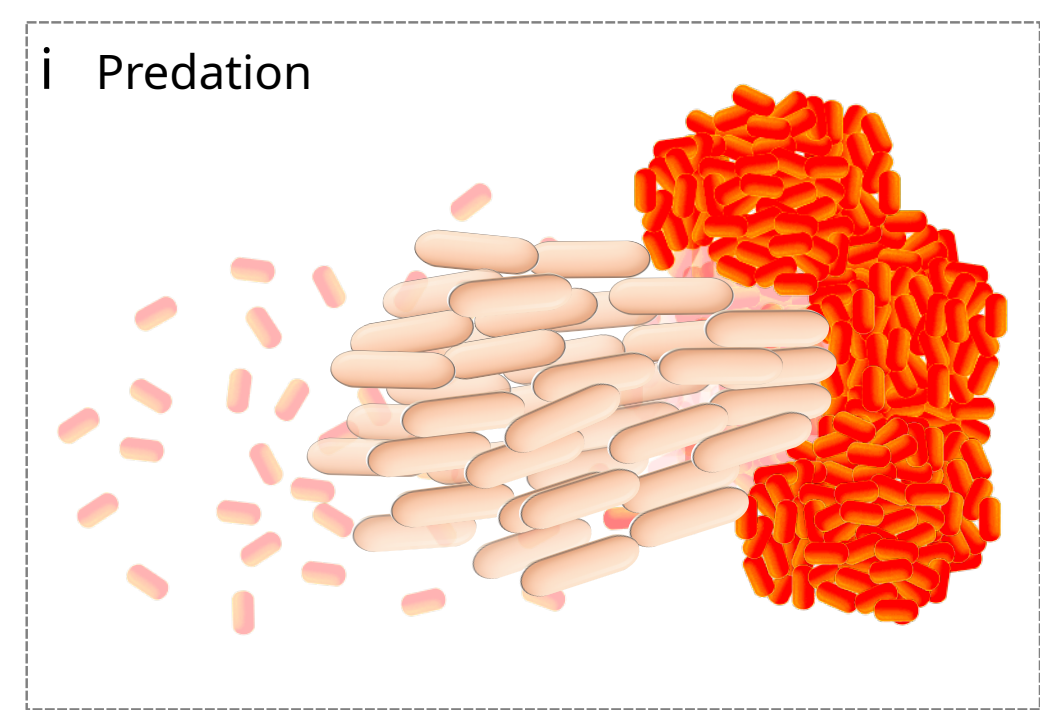

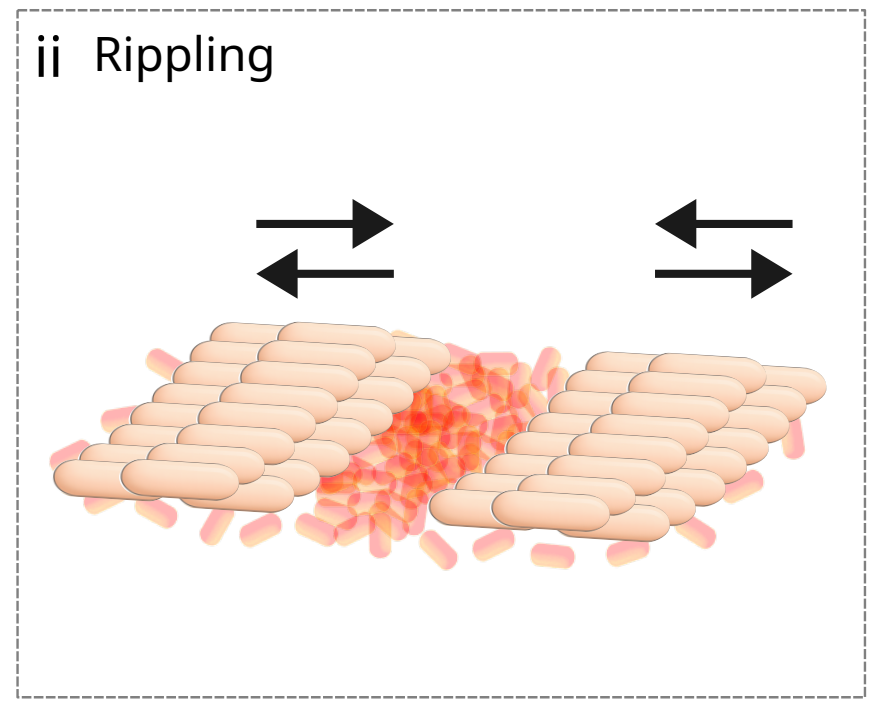

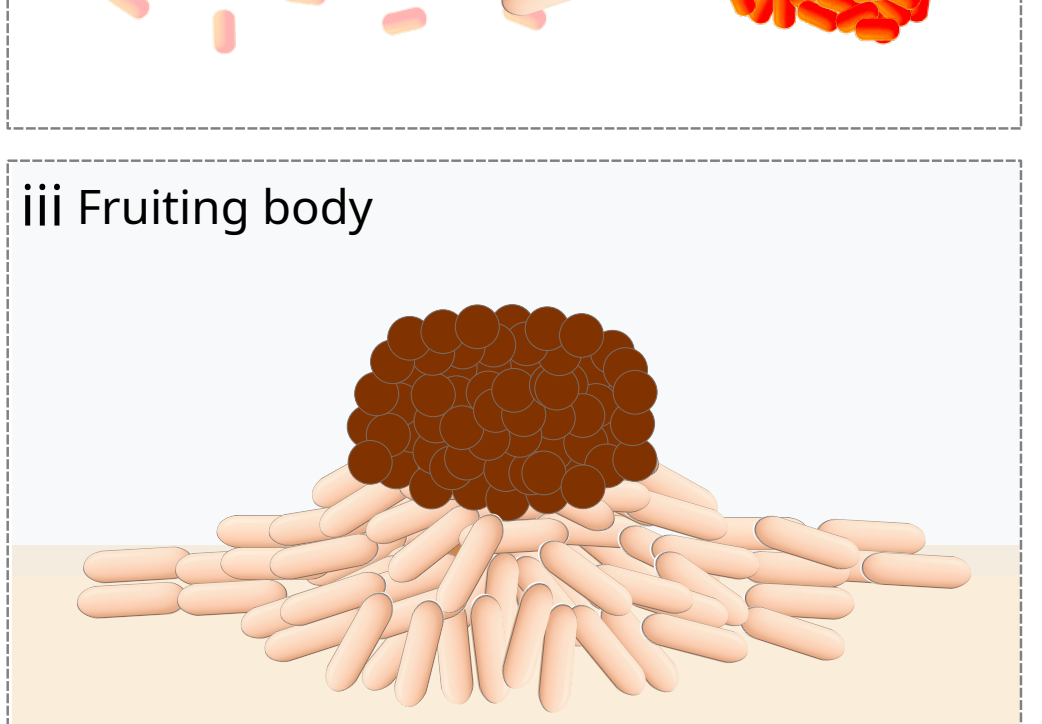

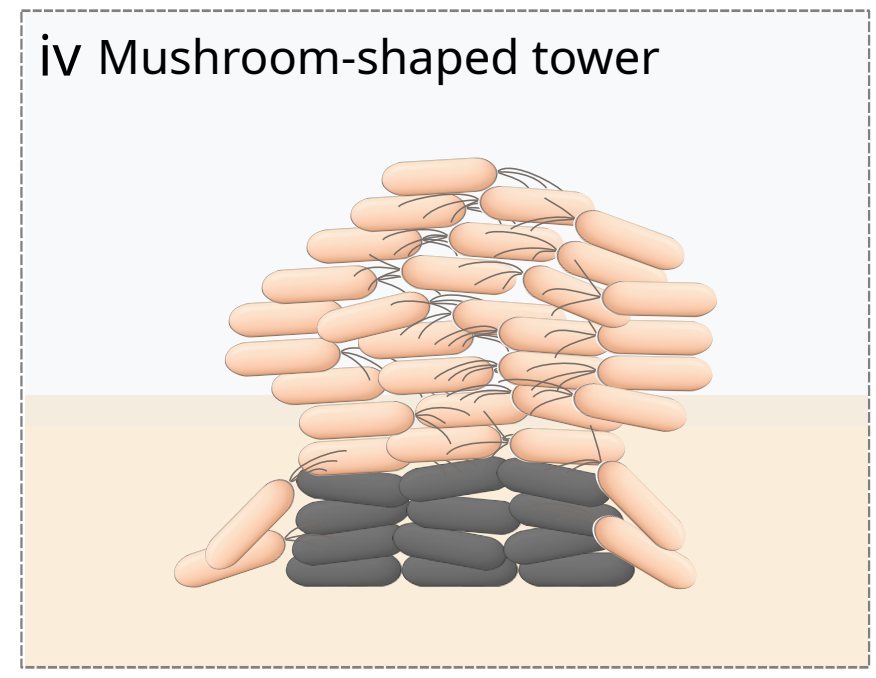

**Figure 2: Type IVa pili-based motility**

**A) Twitching behaviors.** Classical twitching *(i)* involves intermittent attachment-retraction cycles that produce jagged trajectories and reversals; polarity switching relocalizes the T4aP machinery to the opposite pole, reversing movement direction. In *P. aeruginosa (ii)*, simultaneous pili activity can sometimes lift cells into an upright position, with extended pili supporting surface translocation across irregular substrates. Panel *(ii)* inspired by Gibiansky et al.[133]

**B) Collective dynamics of twitching.** (i-iii) *M. xanthus :* (i) Social (S-) motility mediated by T4aP promotes group migration and predation. *M. xanthus* cells are shown in beige, prey cells in orange, and dead prey in light orange. (ii) Ripple patterns emerge during development and predation, promote alignment and signaling, and depend on both A- and S- motility. (iii) Under starvation, vegetative cells (beige) aggregate via A- and S- motility to form multicellular fruiting bodies containing myxospores (brown circles) (iv) *P. aeruginosa*: Twitching motility contributes to the formation of mushroom-shaped towers during biofilm formation; motile cells (beige, with T4aP) actively migrate and climb around non-motile cells (grey). Panels i-iii inspired by Dinet et al.[257]; panel iv inspired by Klausen et al.[22].

**A** **Gliding behaviors**

*(M. xanthus)*

*(F. johnsoniae)*

i

Reversing

t = 0

t = 1

ii

Left-turn bias

**B** **Collective dynamics of gliding**

*(M. xanthus)*

*(F. johnsoniae)*

i

ii

CCW

**Figure 3: Adhesin-based motility - Gliding**
**A) Gliding behaviors.** (i) *M. xanthus*: A-motile cells alternate straight runs and reversals; polarity switching relocates the motility machinery to the opposite pole, reversing movement direction. (ii) *F. johnsoniae*: connected cells under starvation display a left-turn bias Panel ii inspired by Nakane et al.[21].
**B) Collective dynamics of gliding.** (i) *M. xanthus*: A-motility supports exploratory behavior and plays a key role in predation; motile cells penetrate prey colonies, and leave adhesive trails that guide following groups, increasing efficiency of collective attacks. *M. xanthus* cells are shown in beige, prey cells in orange, and dead prey in light orange. (ii) *F. johnsoniae*: Collective vortex formation under nutrient limitation, guided by left-turn bias and extracellular matrix deposition. Panel ii inspired by Nakane et al. [21]

**A** **Label-free microscopy**

Digital holography (DHM)

Interferometric Scattering (iSCAT)

i

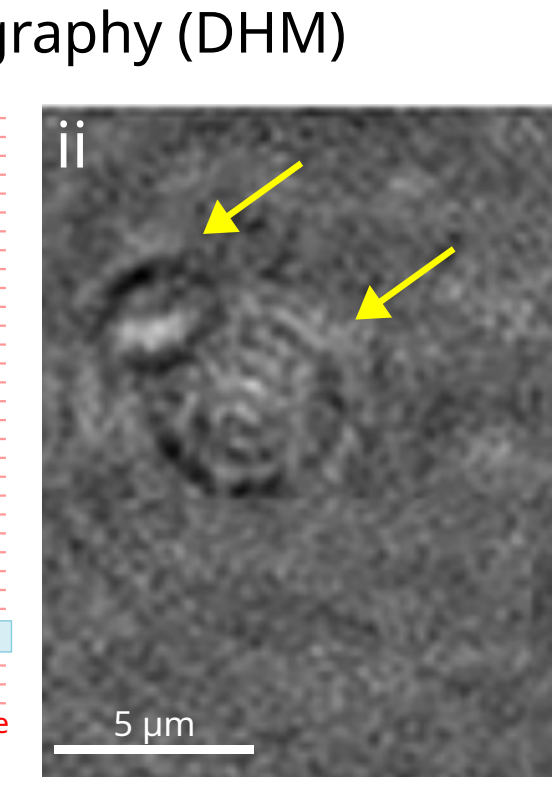

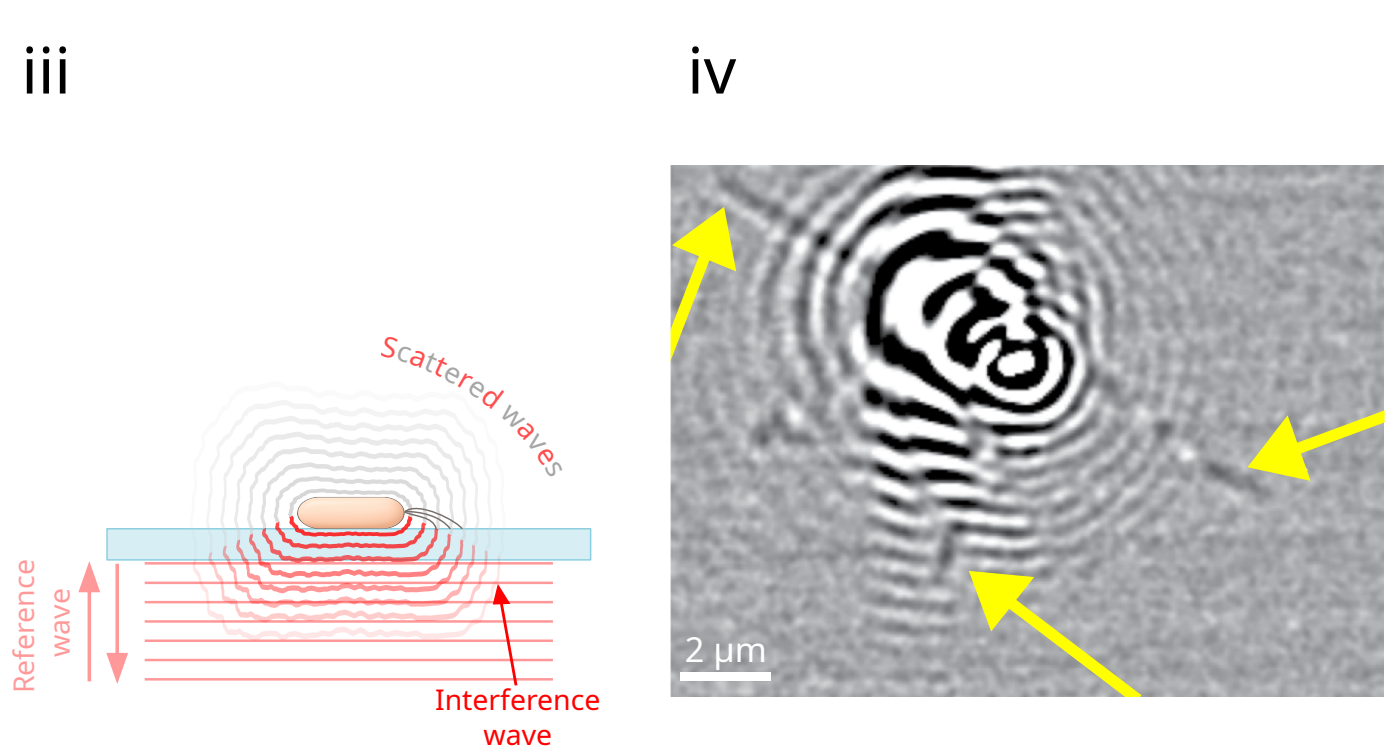

**B** **Widefield illumination**

Epifluorescence

Total internal reflection fluorescence (TIRF)

Highly inclined and laminated optical sheet (HILO)

i

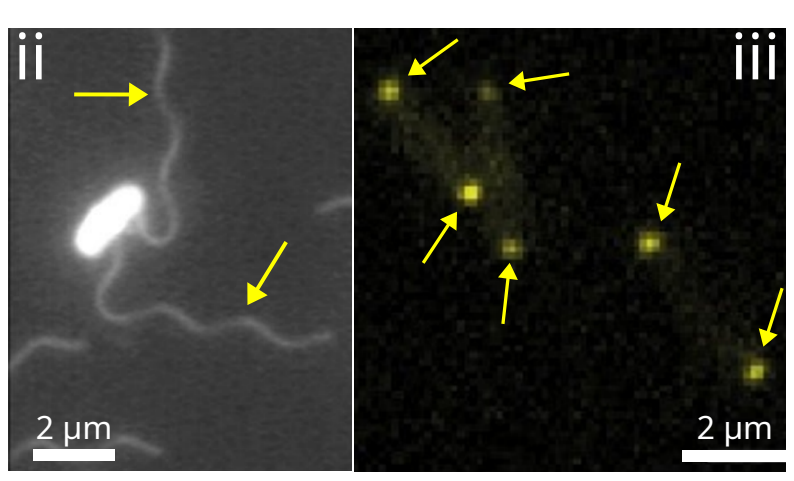

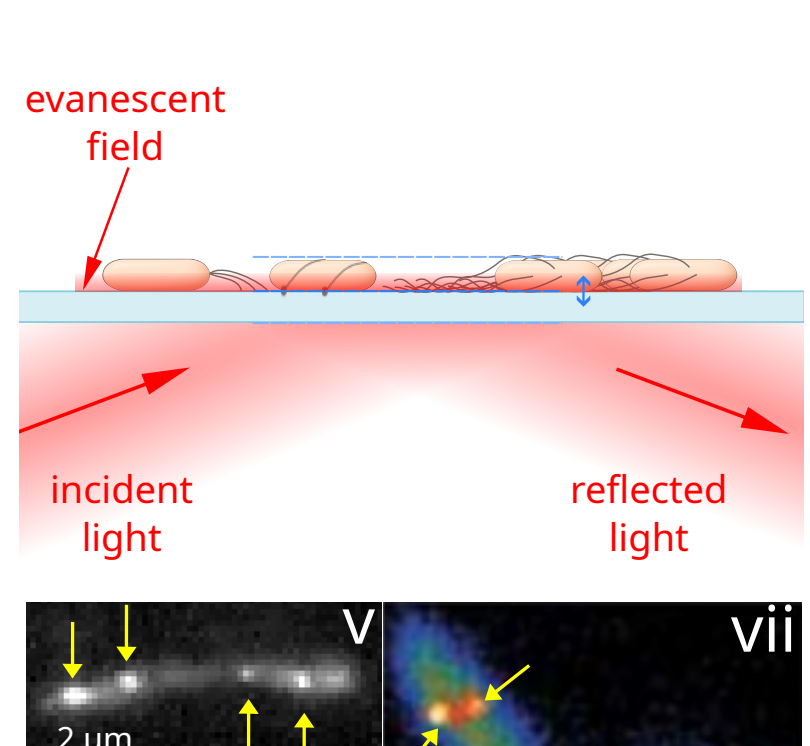

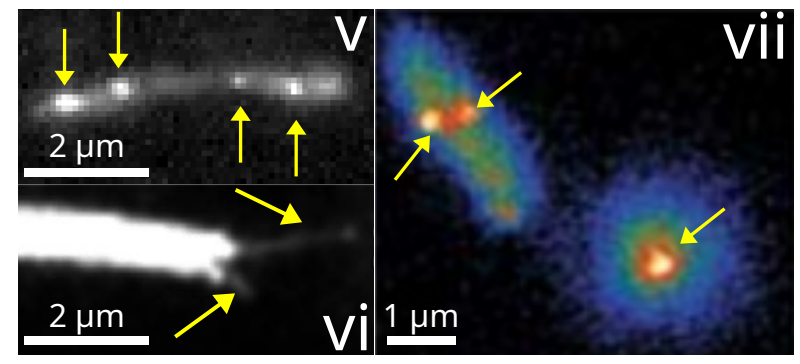

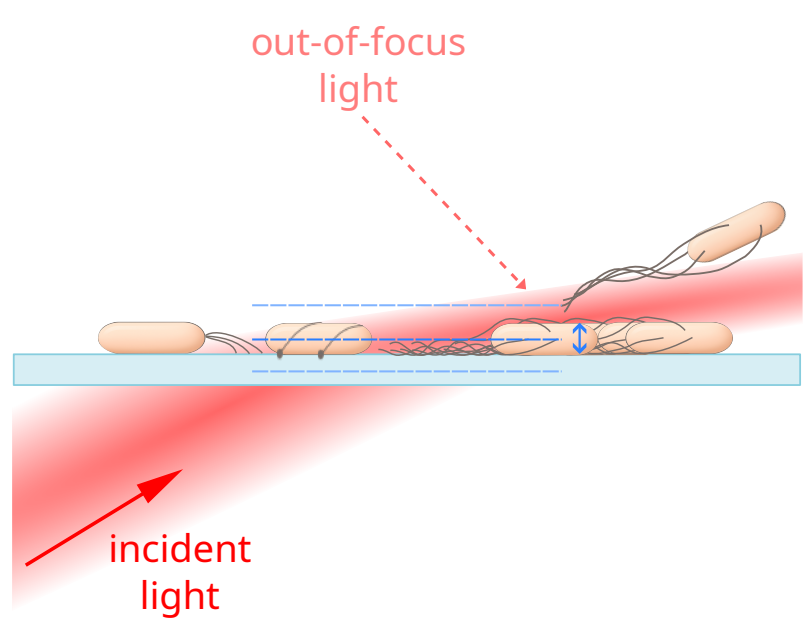

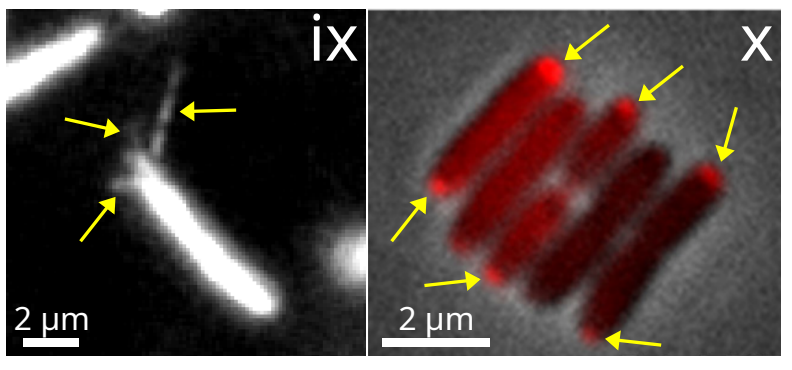

**C** **Point-scanning illumination with optical sectioning**

**D** **Light-sheet illumination with optical sectioning**

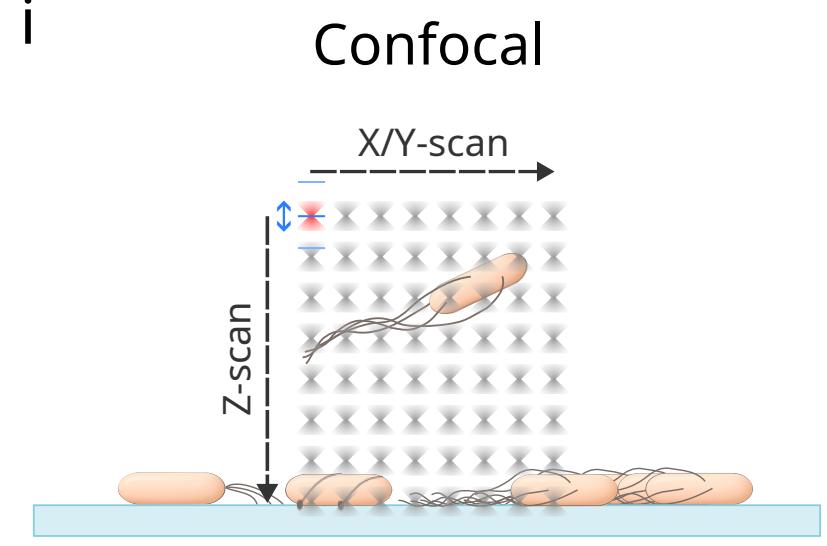

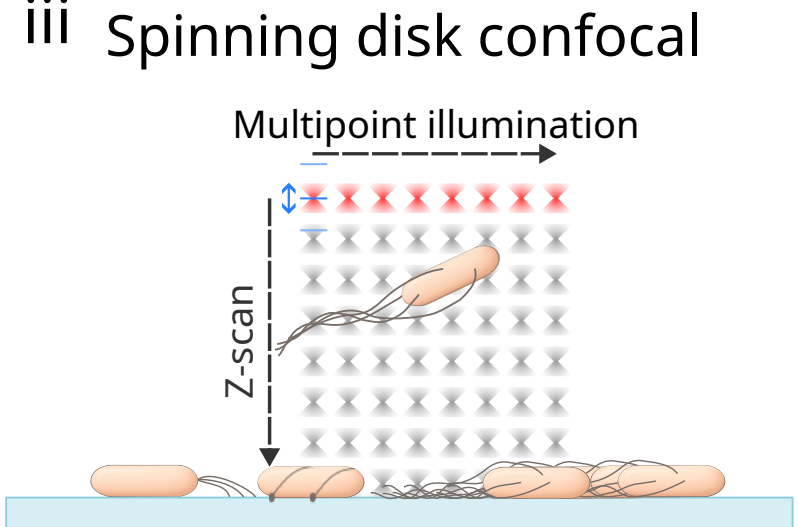

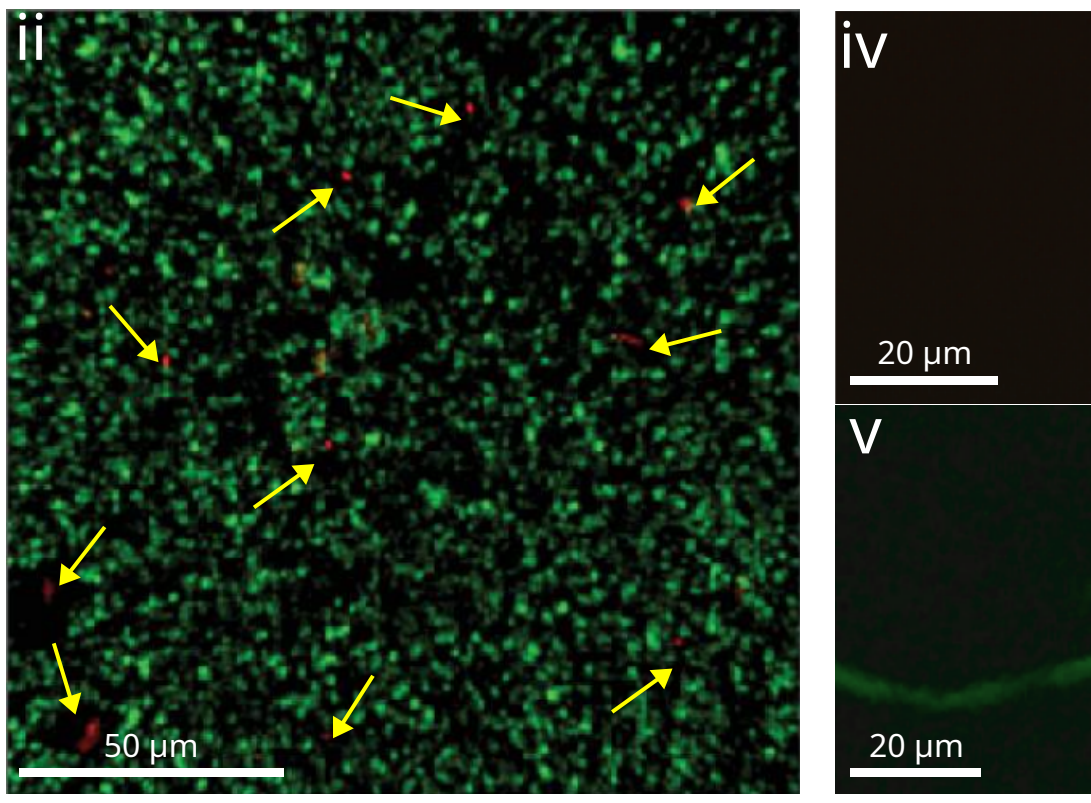

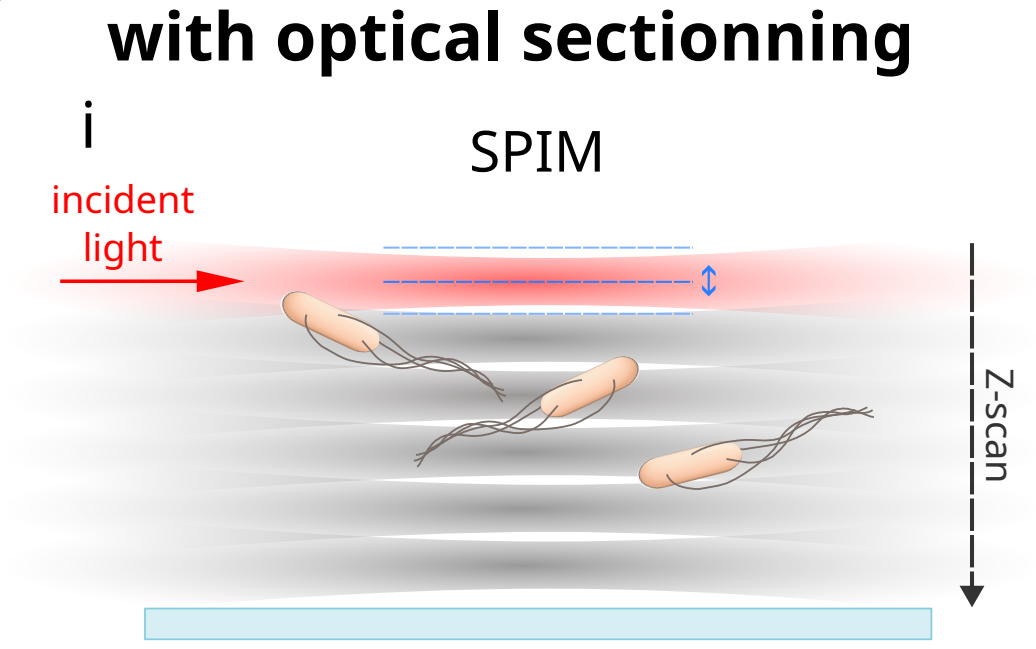

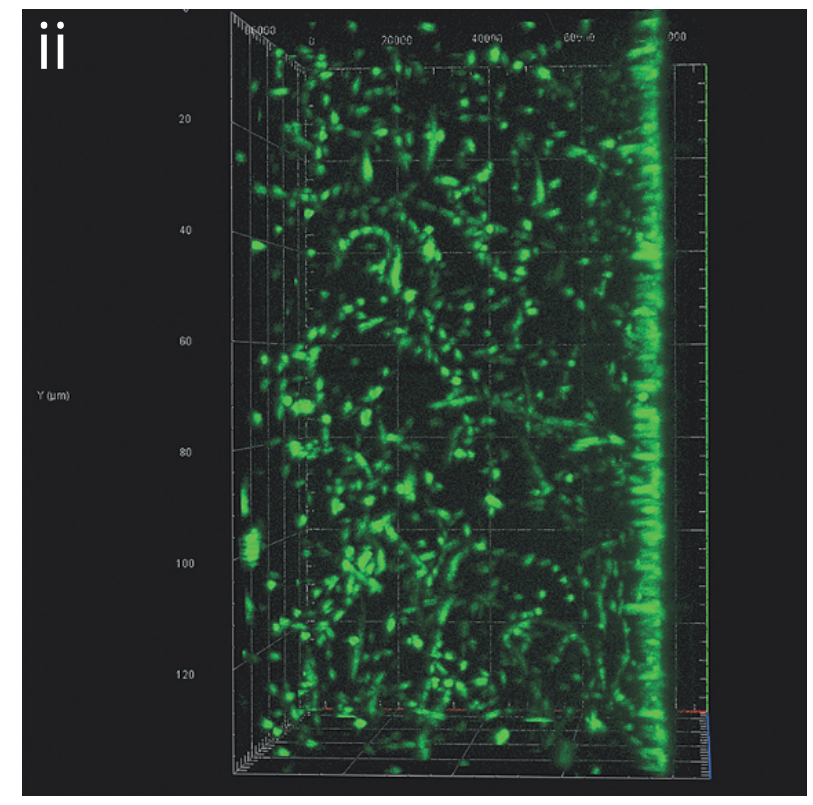

**Figure 4: Microscopy techniques for studying bacterial motility**

**A) Label-free microscopy.** (i) Digital holographic microscopy (DHM) - in-line configuration: Reference wave (red lines) passes through the sample; scattered waves from cells interfere with it to generate a hologram (grey waves scattered upwards are not detected). Phase encodes cell morphology and axial position; numerical reconstruction enables refocusing and 3D tracking of swimming bacteria. (ii) Example DHM amplitude image of *V. alginolyticus* swimming (yellow arrow). Adapted from Nadeau et al.[258]. (iii) Interferometric scattering microscopy (iSCAT): Reference wave from the glass-water reflection interferes with scattered light from nanoscale structures (e.g., pili), producing high-contrast signals with nanometric axial sensitivity. (iv) Example iSCAT image of *P. aeruginosa fliC*$^{-}$ cells showing type IV pili (yellow arrow). Adapted from Talà et al.[120].

**B) Widefield illumination.** (i) Epifluorescence microscopy: Incidence light enters through the objective, illuminates the whole field; emitted fluorescence is collected back through the same objective (fast imaging, but reduced contrast due to out-of-focus signal). (ii-iii) Example epifluorescence image of (ii) *E. coli* with Oregon Green 514-labeled flagella (yellow arrow). Adapted from Turner et al.[51]. (iii) *P. aeruginosa* expressing mNG-PilG (yellow arrow). Adapted from Kühn et al.[136]. (iv) TIRF microscopy: Laser enters the coverslip at an angle exceeding the critical angle, generating an evanescent field (~100-200 nm) that excites only membrane-proximal fluorophores (high surface contrast). (v-vii) Examples of TIRF imaging of (v) AglZ-YFP clusters in *M. xanthus* (yellow arrow, personal data). (vi) Alexa Fluor 488-labeled type IV pili in *M. xanthus* (yellow arrow). Adapted from Mercier et al.[259]. (vii) GFP-MotB in *E. coli* (yellow arrow). Adapted from Leake et al.[194]. (viii) HILO microscopy: Laser at shallow angle below total internal reflection forms a thin inclined illumination field (~1-3 µm) for whole-cell illumination with reduced background. (ix-x) Examples of HILO imaging of (ix) Alexa Fluor 488-labeled type IV pili in *M. xanthus* (yellow arrow, personal data). (x) mRuby3-PilT in *P. aeruginosa* (yellow arrow). Adapted from Koch et al.[196].

**C) Point-scanning illumination with optical sectioning.** (i) Confocal laser scanning microscopy (CLSM): Focused laser excites the sample point by point; fluorescence passes through a pinhole that rejects out-of-focus light, allowing optical sectioning and 3D reconstruction. Acquisition is relatively slow, limiting capture of fast dynamics. (ii) Example of 2D confocal imaging of *Staphylococcus aureus* (green cells, GFP) biofilm and swimming *Bacillus* spp. (red cells, yellow arrow, SYTO61) within it. Adapted from Ravel et al.[197]. (iii) Spinning disk confocal microscopy: A rotating Nipkow disk with thousands of pinholes splits the laser beam into multiple foci, scanning many points in parallel and thus increasing speed while reducing phototoxicity for 3D live imaging. (iv-v) Examples of spinning disk confocal imaging in 2D of (iv) *Borrelia burgdorferi* (yellow arrow, GFP) escaping postcapillary venules in mouse vasculature (Alexa Fluor 555) (v) *B. burgdorferi* (yellow arrow, GFP) imaged in a living mouse ear. Panel iv-v adapted from Moriarty et al.[200].

**D) Planar illumination with optical sectioning.** (i) Light-sheet microscopy (SPIM): Thin illumination sheet excites only the focal plane; detection is orthogonal to illumination, minimizing photobleaching and enabling rapid volumetric imaging of large 3D samples. (ii) Example of SPIM: Reconstructed 3D trajectories of *P. aeruginosa* (DNA-intercalating dye) swimming in agarose chamber. Adapted from Khong et al.[204].

| Technique | Im. mod. | Ph-tox. | Recommended applications | Limitations |
|---|---|---|---|---|
| **Brightfield** | L-F | - | All types of motility; cell body visualization | No specificity; Low contrast |
| **Phase contrast** | L-F | - | All types of motility; cell body visualization; flagella visible if in large bundles | No specificity; Interference halos |
| **DIC** | L-F | - | All types of motility; cell body visualization; flagella visible if in large bundles | No specificity; Directional contrast; visibility depends on orientation |
| **Darkfield** | L-F | - | All types of motility;cell body visualization; flagella visible if in large bundles | No specificity; Highly sensitive to optical noise (dust, bubbles); Incompatible with high-resolution imaging |
| **DHM** | L-F | - | Swimming: single-cell 3D imaging in liquid | No specificity; Phase artifacts, parasitic interferences; reconstruction sensitive to noise; not suitable for dense or scattering media (biofilms, tissues) |
| **iSCAT** | L-F | - | Twitching; label-free visualization of pili; nanometric detection near substrate | No specificity; Limited field of view; sensitive to background noise and surface irregularities; complex interpretation of signal contrast |
| **Epifluoresc-ence** | Fluo. | ++ | All types of motility; visualization of motility reporters; labelled flagella/pili | Out-of-focus fluorescence; poor deep contrast; photobleaching |
| **TIRF** | Fluo. | + | Swarming, twitching and gliding ; visualization of surface-associated reporters; labelled pili visible near substrate | Depends on laser angle; excitation limited to <200 nm above the substrate |
| **HILO** | Fluo. | +/++ | All types of motility; visualization of motility reporters; labelled flagella/pili | Oblique excitation; residual fluorescence out of focal plane |
| **Confocal (CLSM)** | Fluo. | +++ | All types of motility; 3D imaging; suitable for biofilms or tissues; motility reporters, flagella | Slow scanning; unsuitable for fast dynamics |
| **Spinning disk confocal** | Fluo. | ++ | All types of motility; 3D imaging; suitable for biofilms or tissues; motility reporters, flagella | Z-resolution limited in thick samples |
| **Light-sheet (SPIM)** | Fluo. | + | Swimming in liquid media ; swimming/twitching in biofilms or tissues accessible laterally; 3D imaging | Not suitable for surface-adhering bacteria; illumination requires lateral access to sample |

**Table 1 - Comparative overview of microscopy techniques for studying bacterial motility.** Summary of key imaging parameters, advantages, and limitations for label-free and fluorescence-based methods. *L-F, label-free; Fluo., fluorescence; Ph-tox., phototoxicity; Im. mod., imaging modality.*

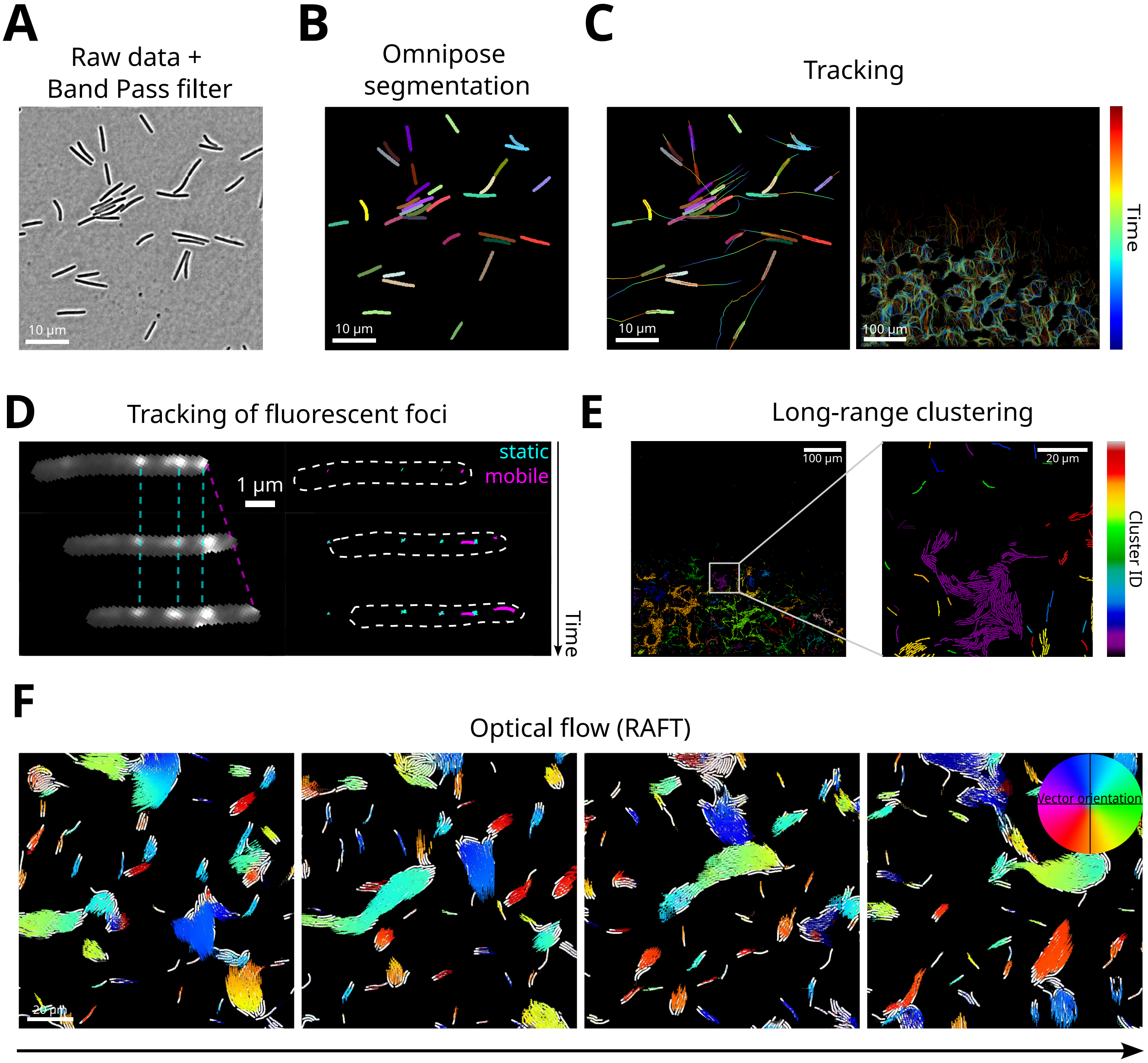

A
Raw data +
Band Pass filter
10 µm
B
Omnipose
segmentation
10 µm
C
Tracking
10 µm
100 µm
Time
D
Tracking of fluorescent foci
1 µm
static
mobile
Time
E
Long-range clustering
100 µm
20 µm
Cluster ID
F
Optical flow (RAFT)
20 µm
Vector orientation
Time

**Figure 5: Image processing and analysis pipelines for bacterial motility**

**A)** Raw phase-contrast microscopy image of *M. xanthus* cells processed with a band-pass filter (personal data).

**B)** Omnipose segmentation of *M. xanthus* cells (A); color-coded by object label (personal data).

**C)** Trajectory reconstruction from cell centroids tracked over 1000 s using the custom pipeline of Mas et al.[132] (left; personal data) and example of *M. xanthus* predation front trajectories after 7 h (right; adapted from Rombouts et al.[12]).

**D)** Tracking of AglZ-YFP fluorescent foci in *M. xanthus* over 150 s, showing the classification of static and mobile clusters, using the custom pipeline of Mas et al.[132] (personal data).

**E)** Long-range clustering analysis of *M. xanthus* cells in close spatial proximity (adapted from Rombouts et al.[12]).

**F)** Optical flow analysis using RAFT[255] applied to probability maps from the segmentation pipeline of Rombouts et al.[218] on *M. xanthus* time-lapse data. Vectors are color-coded by orientation (personal data).